\documentclass[a4paper,11pt]{article}
\usepackage{epsfig}
\usepackage{amssymb}
\usepackage{amsmath}

\begin{document}

\title{Report on the Detailed Calculation of the Effective Potential in Spacetimes with $S^1\times R^d$ Topology and at Finite Temperature}
\author{V.K.Oikonomou\thanks{%
voiko@physics.auth.gr}\\
Dept. of Theoretical Physics Aristotle University of Thessaloniki,\\
Thessaloniki 541 24 Greece} \maketitle
\begin{abstract}
In this paper we review the calculations that are needed to obtain
the bosonic and fermionic effective potential at finite
temperature and volume (at one loop). The calculations at finite
volume correspond to $S^1\times R^d$ topology. These calculations
appear in the calculation of the Casimir energy and of the
effective potential of extra dimensional theories. In the case of
finite volume corrections we impose twisted boundary conditions
and obtain semi-analytic results. We mainly focus in the details
and validity of the results. The zeta function regularization
method is used to regularize the infinite summations. Also the
dimensional regularization method is used in order to renormalize
the UV singularities of the integrations over momentum space. The
approximations and expansions are carried out within the
perturbative limits. After the end of each section we briefly
present applications associated to the calculations. Particularly
the calculation of the effective potential at finite temperature
for the Standard Model fields, the effective potential for warped
and large extra dimensions and the topological mass creation. In
the end we discuss on the convergence and validity of one of the
obtained semi-analytic results.
\end{abstract}

\bigskip

\noindent \textbf{Keywords:} Effective potential, zeta
regularization, Casimir energy, finite temperature, extra
dimensions
\bigskip
\bigskip
\section{Introduction}
During the development of Quantum Field Theory, many quantitative
methods have been developed. Some of the most frequently used
techniques are one-dimensional infinite lattice sums
\cite{elizalde,kirstennew1}. In this article we shall review the
calculations associated with these summations, that appear in many
important branches of Quantum Field Theory, three of which are,
the physics of extra dimensions
\cite{antoniadis,antoniadis2,extra1,extra2,scherk,maclemente,kribs},
the Casimir effect, \cite{elizalde2}-\cite{kirsten14},
\cite{elizalde,gongcharov,elizalde3,ford,spalluci,ferrer,alles}
and finally in field theories at finite temperature
\cite{Das,kapusta,dolan,quirowfinite,elizalde,elizalde2,elizalde3,elizaldesp14,kirstennew1,kirstennew5}.
In both three cases we shall compute the effective potential. The
method we shall use involves the expansion of the potential in
Bessel series and zeta regularization
\cite{elizalde,elizalde2,kirstennew1,elizaldenew1}. We focus on
the details of the calculation and we thing the paper will be a
useful tool for the ones that want to study these theories.
\subsection{Effective Potential in Theories with Large Extra Dimensions}

In theories with large extra dimensions
\cite{antoniadis,antoniadis2,extra1,extra2,scherk,maclemente,kribs},
the fields entering the Lagrangian are expanded in the
eigenfunctions of the extra dimensions. Let us focus on theories
with one extra dimension with the topology of a circle, namely of
the type $S^1\times M_4$ ($M_4$ stands for the 4-dimensional
Minkowski space). In the following we shall also discuss the
orbifold compactification apart from the circle compactification
we describe here. For circle compactifications, the harmonic
expansion of the fields reads,
\begin{equation}
\phi(x,y)=\sum_{n=-\infty }^{\infty
}{\phi}_{n}(x)e^{\frac{i2{\pi}ny}{L}},
\end{equation}
where $x$ stands for the 4-dimensional Minkowski space
coordinates, $y$ for the extra dimension and $L$ the radius of the
extra dimension. We note that fields are periodic in the extra
dimension $y$ namely, $\phi(x,y)=\phi(x,y+2\pi R)$. One of the
ways to break supersymmetry is the Scherk-Schwarz compactification
mechanism. This is based on the introduction of a phase $q$. For
fermions we denote it $q_F$ and for bosons $q_B$. Now the harmonic
expansions for fermion and bosons fields read,
\begin{equation}\label{pordus}
\phi(x,y)=\sum_{n=-\infty }^{\infty
}{\phi}_{n}(x)e^{\frac{i2{\pi}(n+q_{F})y}{L}},
\end{equation}
for fermions and,
\begin{equation}\label{pordius}
\phi(x,y)=\sum_{n=-\infty }^{\infty
}{\phi}_{n}(x)e^{\frac{i2{\pi}(n+q_B)y}{L}},
\end{equation}
for bosons. We can observe that the initial periodicity condition
is changed. Using equations (\ref{pordus}) and (\ref{pordius}) we
can find that the effective potential at one loop is equal to,
\begin{equation}\label{extrdimens}
V(\phi)=\frac{1}{2}\mathrm{Tr}\sum_{n=-\infty }^{\infty }\int
\frac{\mathrm{d^4p}}{(2\pi)^4}\ln
\Big{[}\frac{p^2+\frac{(n+q_B)^2}{L^2}+M^2(\phi)}{{p^2+\frac{(n+q_F)^2}{L^2}+M^2(\phi)}}\Big{]}
\end{equation}
Note that fermions and bosons contribute to the effective
potential with opposite signs. This is due to the fact that
fermions are described by anti-commuting Grassmann fields. Also
$M^2(\phi)$ is a $n$ independent term and depends on the way that
spontaneous symmetry breaking occurs. We shall not care for the
particular form of this and we focus on the general calculation of
terms like the one in equation (\ref{extrdimens}).

\subsection{The Casimir Energy}
One of the most interesting phenomena in Quantum Field Theory is
the Casimir effect (for a review see
\cite{elizalde,elizalde2,bordagreview,direstr1,kirstennew1,Milton1}).
It expresses the quantum fluctuations of the vacuum of a quantum
field. It originates from the "confinement" of a field in finite
volume. Many studies have been done since H. Casimir's original
work \cite{Casimir}. The Casimir energy, usually calculated in
these studies, is closely related to the boundary conditions of
the fields under consideration
\cite{elizaldesp24,elizaldesp27,elizaldesp1,elizaldesp12,elizalde,elizalde2,odi,odi1}.
Boundary conditions influence the nature of the so-called Casimir
force, which is generated from the vacuum energy.

\noindent In this paper we shall concentrate on the computation of
the effective potential (Casimir Energy) of bosonic and fermionic
fields in a space time with the topology $S^1\times R^d$
\cite{elizalde,elizalde2,bordagreview,elizaldesp19,elizaldesp23,elizaldesp25,elizaldesp26,kirstennew1}.
Fermionic and bosonic fields in spaces with non trivial topology
are allowed to be either periodic or anti-periodic in the compact
dimension. The forms of the potential to be studied are,
\begin{equation}\label{1071}
\frac{1}{L}\int \frac{dk^{d}}{(2\pi )^{d}}\sum_{n=-\infty }^{\infty }\ln [%
\frac{4\pi^{2} n^{2}}{L^{2}}+k^{2}+m^{2}] ,\end{equation} and the
fermionic one,
\begin{equation}\label{1081}
\frac{1}{L}\int \frac{dk^{d}}{(2\pi )^{d}}\sum_{n=-\infty }^{\infty }\ln [%
\frac{(2n+1)^{2}\pi^{2}}{L^{2}}+k^{2}+m^{2}] .\end{equation} We
shall study them also in the cases $d=2$ and $d=3$, which are of
particular importance in physics since they correspond to three
and four total dimensions. Both have many applications in solid
state physics and cosmology \cite{bordagreview,elizalde}. Also we
shall generalize to the case with fermions and bosons obeying
general boundary conditions also in $d+1$ dimensions. This is
identical from a calculational aspect with the effective potential
of theories with extra dimensions \cite{maclemente,extra1}. So
computing one of the two gives simultaneously the other. The
expression that is going to be studied thoroughly is,
\begin{align}\label{1341}
&\frac{1}{L}\int \frac{dk^{d}}{(2\pi )^{d}}\sum \ln [((n+\omega
)\frac{2\pi
}{L})^{2}+k^{2}+m^{2}]=\\&\notag \int \frac{dk^{d+1}}{(2\pi )^{d+1}}%
\ln [k^{2}+a^{2}]\\&\notag +\frac{1}{L}\int \frac{dk^{d}}{(2\pi
)^{d}}\ln [1-e^{-2(\frac{aL}{2}-i\pi
\omega )}]+\frac{1}{L}\int \frac{dk^{d}}{(2\pi )^{d}}\ln [1-e^{-2(\frac{aL}{2%
}+i\pi \omega )}] .\end{align} The calculations shall be done in
$d+1$ dimensions, quite general, and the application to every
dimension we wish, can be done easily. The only constraint shall
be if $d$ is even or odd. We shall make that clear in the
corresponding sections and treat both cases in detail.

\subsection{Field Theories at Finite Temperature}

The calculations used in finite temperature field theories are
based on the imaginary time formalism
\cite{kapusta,Das,elizalde,kirstennew1,elizalde2}:
\begin{equation}\label{1}
t\rightarrow i\beta ,\end{equation} with $\beta $=$\frac{1}{T}$.
The eigenfrequencies of the fields that appear to the propagators
are discrete and are summed in the partition function. These are
affected from the boundary conditions used for fermions and bosons
\cite{elizalde,elizalde2}. Bosons obey only periodic and fermions
antiperiodic boundary conditions at finite temperature, as we
shall see (this is restricted and dictated by the KMS relations
\cite{Das}). Indeed for bosons the boundary conditions are:
\begin{equation}\label{2}
\varphi (x,0)=\varphi (x,\beta ) ,\end{equation} where $x$ stands
for space coordinates, and the fermionic boundary conditions are,
\begin{equation}\label{3}
\psi (x,0)=-\psi (x,\beta ) ,\end{equation} In most calculations
involving bosons, we are confronted with the following expression:
\begin{equation}\label{4}
T\int \frac{dk^{3}}{(2\pi )^{3}}\sum_{n=-\infty }^{\infty
}\ln[4\pi^{2} n^{2}T^{2}+k^{2}+m^{2}] ,\end{equation} while the
fermionic contribution is,
\begin{equation}\label{5}
T\int \frac{dk^{3}}{(2\pi )^{3}}\sum_{n=-\infty }^{\infty
}\ln[(2n+1)^{2}\pi^{2}T^{2}+k^{2}+m^{2}] ,\end{equation} and $k$
stands for the Euclidean momentum:
\begin{equation}\label{6}
k^{2}=k_{1}^{2}+k_{2}^{2}+k_{3}^{2} ,\end{equation} while $m$ is
the field mass. In the next sections we deal with the two above
contributions in $d+1$ dimensions and we specify the results for
$d=3$ and $d=2$.

\section{Bosonic Contribution at Finite Temperature}

We will compute the following expression,
\begin{equation}\label{7}
S_1=T\int \frac{dk^{3}}{(2\pi )^{3}}\sum_{n=-\infty }^{\infty
}\ln[4\pi^{2} n^{2}T^{2}+k^{2}+m^{2}]
\end{equation}
In the following we generalize in $d$ dimensions. This will give
us the opportunity to deal other cases apart from the $d=4$.
Consider the sum:
\begin{equation}\label{8}
S_o=\sum\limits_{n=-\infty }^{\infty }\frac{1}{4\pi^{2}n^{2}T^{2}+a^{2}}=\frac{%
1}{4\pi^{2}T^{2}}\sum\limits_{n=-\infty }^{\infty }\frac{1}{n^{2}+\frac{%
a^{2}}{4\pi^{2}T^{2}}} ,\end{equation} where,
\begin{equation}\label{9}
a^{2}=k^{2}+m^{2}
\end{equation}
Integrating over $a^{2}$,
\begin{equation}\label{10}
\sum\limits_{n=-\infty }^{\infty
}\frac{1}{4\pi^{2}n^{2}T^{2}+a^{2}} ,\end{equation} we get:
\begin{equation}\label{11}
\ \int \ \sum\limits_{n=-\infty }^{\infty
}\frac{da^{2}}{4\pi^{2}n^{2}T^{2}+a^{2}}=\sum
\ln[4\pi^{2}n^{2}T^{2}+a^{2}] .\end{equation} Now,
\begin{equation}\label{12}
\sum\limits_{n=-\infty }^{\infty }\frac{1}{4\pi^{2}n^{2}T^{2}+a^{2}}=%
\frac{2}{4aT}\coth (\frac{a}{2T}) ,\end{equation} thus equation
(\ref{11}) becomes,
\begin{eqnarray}\label{13}
\ \int \ \sum\limits_{n=-\infty }^{\infty
}\frac{da^{2}}{4\pi^{2}n^{2}T^{2}+a^{2}} &=&\int \frac{2}{4aT}(\coth (\frac{a}{2T})da^{2} \\
&=&2\ln(\sinh [\frac{a}{2T}]) \nonumber .\end{eqnarray} Using the
relation \cite{qradstein7},
\begin{equation}\label{14}
\ln(\sinh
x)=\ln(\frac{1}{2}[e^{x}-e^{-x}])=x+\ln[1-e^{-2x}]-\ln[2]
,\end{equation} and upon summation,
\begin{equation}\label{15}
\ln(\sinh\frac{a}{2T})=\frac{a}{2T}+\ln[1-e^{-\frac{a}{T}}]-\ln[2]
,\end{equation} and,
\begin{equation}\label{16}
\ln (\sinh \frac{a}{2T})=\frac{a}{2T}+\ln [1-e^{-\frac{a}{T}}]-\ln
[2] .\end{equation} Summing equations (\ref{15}) and (\ref{16}) we
obtain,
\begin{equation}\label{17}
\int \sum_{n=-\infty }^{\infty
}\frac{da^{2}}{4\pi^{2}n^{2}T^{2}+a^{2}}=2\ln (\sinh
\frac{a}{2T})=\frac{a}{T} +2\ln [1-e^{-\frac{a}{T}}]-2\ln [2]
.\end{equation} Finally the result is
\cite{kapusta,Das,elizalde,kirstennew1}:
\begin{equation}\label{18}
\sum\limits_{n=-\infty }^{\infty }\ln
[4\pi^{2}n^{2}T^{2}+a^{2}]=\frac{a}{T}+2\ln
[1-e^{-\frac{a}{T}}]-2\ln [2]\ .\end{equation} Upon using,
\begin{equation}\label{20}
\sum \ln [\frac{(n+\omega )^{2}4\pi^{2}T^{2}+a^{2})}{(n+\omega
)^{2}4\pi^{2}T^{2}+b^{2})}]=2(a-b) ,\end{equation} equation
(\ref{18}) becomes,
\begin{equation}\label{21}
\sum \ln [4\pi^{2}n^{2}T^{2}+a^{2}]=\frac{1}{2\pi
T}\int\limits_{-\infty }^{\infty }dx\ln [x^{2}+a^{2}]+2\ln
[1-e^{-\frac{a}{T}}] .\end{equation} Finally we have,
\begin{align}\label{ape1}
\ T\int \frac{dk^{3}}{(2\pi )^{3}}\sum \ln [(2\pi n
T)^{2}+k^{2}+m^{2}]&=\int \frac{dk^{3}}{(2\pi
)^{3}}\int\limits_{-\infty }^{\infty }\frac{dx}{2\pi }\ln
[x^{2}+a^{2}]\\& \notag +2T\int \frac{dk^{3}}{(2\pi )^{3}}\ln
[1-e^{-\frac{a}{T}}] .\end{align} Remembering that,
\begin{equation}\label{22}
\ a^{2}=k^{2}+m^{2} ,\end{equation} the first integral of equation
(\ref{ape1}) is the one loop contribution to the effective
potential at zero temperature. The 4-momentum is:
\begin{equation}\label{23}
K^{2}=k^{2}+x^{2} .\end{equation} Writing the above in $d+1$
dimensions (in the end we take $d=3$ to come back to four
dimensions) we get,
\begin{align}\label{24}
T\int \frac{dk^{d}}{(2\pi )^{d}}\sum \ln
[4\pi^{2}n^{2}T^{2}+k^{2}+m^{2}]&=\int \frac{dk^{d+1}}{(2\pi
)^{d+1}}\ln [k^{2}+a^{2}] \\& \notag +2T\int \frac{dk^{d}}{(2\pi
)^{d}}\ln [1-e^{-\frac{a}{T}}] .\end{align} The temperature
dependent part has singularities stemming from the infinite
summations. These singularities are poles of the form
\cite{elizalde,kirstennew1,elizalde2}:
\begin{equation}\label{polepock}
    \frac{1}{\epsilon}
,\end{equation} where $\epsilon \rightarrow 0$ the dimensional
regularization variable ($d=4+\epsilon$). As we shall see, by
using the zeta regularization
\cite{elizalde,elizalde2,elizalde3,kirstennew1,elizaldenew1} these
will be erased. In the following of this section we focus on the
calculation of the temperature dependent part. Let,
\begin{equation}\label{25}
V_{boson}=2T\int \frac{dk^{d}}{(2\pi )^{d}}\ln
[1-e^{-\frac{a}{T}}]=2T\int \frac{dk^{d}}{(2\pi )^{d}}\ln
[1-e^{-\frac{a}{T}}] .\end{equation} By using \cite{qradstein7},
\begin{equation}\label{26}
\ln [1-e^{-\frac{a}{T}}]=-\sum\limits_{q=1}^{\infty }\frac{e^{-\frac{a}{T}%
q}}{q} .\end{equation} we obtain,
\begin{eqnarray}\label{28}
V_{boson} &=&2T\int \frac{dk^{d}}{(2\pi )^{d}}\ln
[1-e^{-\frac{a}{T}}]=-2T\int
\frac{dk^{d}}{(2\pi )^{d}}\sum\limits_{q=1}^{\infty }\frac{e^{-\frac{a}{T}q}%
}{q} \\
&=&-2\sum\limits_{q=1}^{\infty }T\int \frac{dk^{d}}{(2\pi )^{d}}\frac{e^{-%
\frac{a}{T}q}}{q}  \nonumber
\end{eqnarray}
and remembering,
\begin{equation}\label{29}
a=\sqrt{k^{2}+m^{2}} ,\end{equation} by integrating over the
angles we get,
\begin{align}\label{30}
V_{boson} &=-2\sum\limits_{q=1}^{\infty }T\int \frac{dk^{d}}{(2\pi
)^{d}}\frac{e^{-\frac{\sqrt{k^{2}+m^{2}}}{T}q}}{q}\\ & \notag
=-2\sum\limits_{q=1}^{\infty }T\int_{-\infty }^{\infty
}\frac{dk}{(2\pi )^{d}}k^{d-1}\frac{(2\pi )^{\frac{d}{2}}}{\Gamma
(\frac{d}{2})}\frac{e^{-\frac{\sqrt{k^{2}+m^{2}}}{T}q}}{q} \\
& \notag=-2\sum\limits_{q=1}^{\infty }T\frac{(2\pi
)^{\frac{d}{2}}}{\Gamma (\frac{d}{2})q(2\pi )^{d}}\int_{-\infty
}^{\infty }dkk^{d-1}e^{-\frac{\sqrt{k^{2}+m^{2}}}{T}q}
.\end{align} The integral,
\begin{equation}\label{31}
\int_{-\infty }^{\infty
}dkk^{d-1}e^{-\frac{\sqrt{k^{2}+m^{2}}}{T}q}\
,\end{equation}%
equals to \cite{qradstein7},
\begin{equation}\label{32}
\int_{-\infty }^{\infty }dkk^{d-1}e^{-\frac{\sqrt{k^{2}+m^{2}}}{T}q}=2^{%
\frac{d}{2}-1}(\sqrt{\pi })^{-1}(\frac{q}{T})^{\frac{1}{2}-\frac{d}{2}}m^{%
\frac{d+1}{2}}\Gamma (\frac{d}{2})K_{\frac{d+1}{2}}(\frac{mq}{T})
.\end{equation} So $V_{boson}$ can be written:
\begin{eqnarray}\label{33}
V_{boson} &=&-2\sum\limits_{q=1}^{\infty }\frac{2^{\frac{d}{2}-1}}{(2\pi )^{d}}%
(2\pi )^{\frac{d+1}{2}}m^{d+1}K_{\frac{d+1}{2}}(\frac{mq}{T})(\frac{T}{mq%
})^{\frac{d+1}{2}} \\
&=&-\sum\limits_{q=1}^{\infty }\frac{1}{(2\pi )^{d}}(2\pi )^{\frac{d-1}{2}%
}m^{d+1}\frac{K_{\frac{d+1}{2}}(\frac{mq}{T})}{(\frac{mq}{2T})^{\frac{d+1}{2}}}
\nonumber .\end{eqnarray} The function \cite{qradstein7},
\begin{equation}\label{34}
\frac{K_{\nu }(z)}{(\frac{z}{2})^{\nu }}=\frac{1}{2}\int\limits_{0}^{\infty }%
\frac{e^{-t-\frac{z^{2}}{4t}}}{t^{\nu +1}}dt\
,\end{equation}%
is even under the transformation $z\rightarrow-z$. Thus equation
(\ref{33}) becomes:
\begin{align}\label{35}
V_{boson}&=-\sum\limits_{q=1}^{\infty }\frac{1}{(2\pi )^{d}}(2\pi
)^{\frac{d-1}{2}}m^{d+1}\frac{K_{\frac{d+1}{2}}(\frac{mq}{T})}{(\frac{mq}{2T})^{\frac{d+1}{2}}}\\&
\notag =-\frac{1}{2}\sum\limits_{q=-\infty }^{\infty \prime
}\frac{1}{(2\pi )^{d}}(2\pi
)^{\frac{d-1}{2}}m^{d+1}\frac{K_{\frac{d+1}{2}}(\frac{mq}{T})}{(\frac{mq}{2T})^{\frac{d+1}{2}}}
.\end{align} (The symbol $ ^{\prime}$ in the summation denotes
omission of the zero mode term $q=0$). By using,
\begin{equation}\label{36}
\frac{K_{\nu }(z)}{(\frac{z}{2})^{\nu }}=\frac{1}{2}\int\limits_{0}^{\infty }%
\frac{e^{-t-\frac{z^{2}}{4t}}}{t^{\nu +1}}dt,\end{equation} we
get,
\begin{equation}\label{37}
V_{boson}=-\frac{1}{4}\frac{1}{(2\pi )^{d}}(2\pi )^{\frac{d-1}{2}%
}m^{d+1}\int_{0}^{\infty }dte^{-t}\frac{\sum\limits_{q=-\infty
}^{\infty
\prime}e^{-\frac{(\frac{mq}{T})^{2}}{4t}}}{t^{\frac{d+1}{2}+1}}
.\end{equation} Let $\lambda =\frac{(\frac{m}{T})^{2}}{4t}$. Using
the Poisson summation formula
\cite{elizaldenew1,kirstennew1,elizalde,elizalde2} we have,
\begin{equation}\label{38}
\sum_{q=-\infty }^{\infty }e^{-\lambda q^{2}}=\sqrt{\frac{\pi }{\lambda }}%
\sum_{k=-\infty }^{\infty }e^{-\frac{4\pi^{2}k^{2}}{4\lambda }}
,\end{equation} and omitting the zero modes we obtain:
\begin{equation}\label{39}
1+\sum_{q=-\infty }^{\infty \prime}e^{-\lambda
q^{2}}=\sqrt{\frac{\pi }{\lambda }} (\ \ 1\ +\
\sum_{k=-\infty}^{\infty \prime}e^{-\frac{4\pi^{2}k^{2}}{4\lambda
}}) .\end{equation} Finally,
\begin{align}\label{40}
\sum_{q=-\infty }^{\infty \prime}e^{-\lambda q^{2}}= \sqrt{\frac{\pi }{\lambda }}%
(\ 1\ +\ \sum_{k=-\infty }^{\infty
\prime}e^{-\frac{4\pi^{2}k^{2}}{4\lambda }})\ -\ 1 ,\end{align}
and replacing in $V_{boson}$ we take,
\begin{align}\label{41}
&V_{boson}=\\&\notag -\frac{1}{4}\frac{1}{(2\pi )^{d}}(2\pi )^{\frac{d-1}{2}%
}m^{d+1}\int_{0}^{\infty }dte^{-t}(\frac{\sqrt{\frac{\pi }{\lambda
}}(\ 1\
+\ \sum_{k=-\infty}^{\infty \prime}e^{-\frac{4\pi^{2}k^{2}}{4\lambda }})\ -\ 1}{%
t^{\frac{d+1}{2}+1}}) .\end{align} Set,
\begin{equation}\label{42}
\ \nu =\frac{d+1}{2} .\end{equation} and equation (\ref{41})
reads,
\begin{eqnarray}\label{43}
V_{boson} &=&-\frac{1}{4}\frac{1}{(2\pi )^{d}}(2\pi )^{\frac{d-1}{2}%
}m^{d+1}(\int_{0}^{\infty }dte^{-t}\frac{\sqrt{\frac{\pi }{\lambda }}\ }{%
t^{\nu +1}})- \\
&&\frac{1}{4}\frac{1}{(2\pi )^{d}}(2\pi )^{\frac{d-1}{2}}m^{d+1}\int_{0}^{%
\infty }dte^{-t}(\frac{\sqrt{\frac{\pi }{\lambda }}(\
\sum_{k=-\infty}^{\infty \prime}e^{-\frac{4\pi^{2}k^{2}}{4\lambda }})\ }{t^{\nu +1}})\nonumber \\
&&+\frac{1}{4}\frac{1}{(2\pi )^{d}}(2\pi )^{\frac{d-1}{2}}m^{d+1}\int_{0}^{%
\infty }dte^{-t}(\frac{1}{t^{\nu +1}})  \nonumber .\end{eqnarray}
Also by setting,
\begin{equation}\label{44}
a=\frac{m}{T},
\end{equation}
equation (\ref{43}) becomes (with $\lambda =\frac{a^{2}}{4t}$),
\begin{align}\label{45}
V_{boson}&=-\frac{1}{4}\frac{1}{(2\pi )^{d}}(2\pi
)^{\frac{d-1}{2}}m^{d+1}(\int_{0}^{\infty }dte^{-t}\frac{\sqrt{\pi
t}2}{at^{\nu +1}})\\&\notag -\frac{1}{4}\frac{1}{(2\pi )^{d}}(2\pi
)^{\frac{d-1}{2}}m^{d+1}\int_{0}^{\infty }dte^{-t}(\frac{\sqrt{\pi
t}2(\ \sum_{k=-\infty}^{\infty
\prime}e^{-\frac{4\pi^{2}k^{2}}{a^{2}}t})\ }{at^{\nu
+1}})\\&\notag +\frac{1}{4}\frac{1}{(2\pi )^{d}}(2\pi
)^{\frac{d-1}{2}}m^{d+1}\int_{0}^{\infty }dte^{-t}(\frac{1}{t^{\nu
+1}}) .\end{align} From this, after some calculations we obtain:
\begin{align}\label{46}
V_{boson}&=-\frac{1}{2}\frac{\sqrt{\pi }}{(2\pi )^{d}a}(2\pi
)^{\frac{d-1}{2}}m^{d+1}(\int_{0}^{\infty }dte^{-t}\ t^{-\nu
-\frac{1}{2}})\\&\notag -\frac{1}{2}\frac{\sqrt{\pi }}{(2\pi
)^{d}a}(2\pi )^{\frac{d-1}{2}}m^{d+1}\int_{0}^{\infty
}dte^{-t}(\frac{\sqrt{\pi t}2(\sum_{k=-\infty}^{\infty
\prime}e^{-\frac{4\pi^{2}k^{2}}{a^{2}}t})\ }{at^{\nu
+\frac{1}{2}}})
\\&\notag +\frac{1}{4}\frac{1}{(2\pi )^{d}}(2\pi
)^{\frac{d-1}{2}}m^{d+1}\int_{0}^{\infty }dte^{-t}(\frac{1}{t^{\nu
+1}}) .\end{align} By using \cite{qradstein7},
\begin{equation}\label{47}
\ \frac{1}{(x^{2}+a^{2})^{\mu +1}}=\frac{1}{\Gamma (\mu
+1)}\int_{0}^{\infty }dte^{-(x^{2}+a^{2})t}t^{\mu }
,\end{equation} we finally have:
\begin{eqnarray}\label{48}
V_{boson} &=&-\frac{1}{2}\frac{\sqrt{\pi }}{(2\pi )^{d}a}(2\pi )^{\frac{d-1}{2}%
}m^{d+1}\Gamma (-\nu -\frac{1}{2}+1)\nonumber \\
&&-\frac{1}{2}\frac{\sqrt{\pi }}{(2\pi )^{d}a}(2\pi )^{\frac{d-1}{2}%
}m^{d+1}\Gamma (-\nu -\frac{1}{2}+1)\nonumber\\
&&\times [\sum_{k=-\infty}^{\infty \prime}(1+(\frac{2\pi k}{a})^{2})^{\nu +\frac{1}{2}-1}]\nonumber\\
&& + \frac{1}{4}\frac{1}{(2\pi )^{d}}(2\pi )^{\frac{d-1}{2}%
}m^{d+1}\Gamma (-\nu ) .\end{eqnarray} The sum,
\begin{equation}\label{49}
\sum_{k=-\infty}^{\infty \prime}(1+(\frac{2\pi k}{a})^{2})^{\nu
+\frac{1}{2}-1} ,\end{equation} is invariant under the
transformation $k\rightarrow -k$. Thus we change the summation to,
\begin{equation}\label{50}
2\sum_{k=1}^{\infty }(1+(\frac{2\pi k}{a})^{2})^{\nu
+\frac{1}{2}-1} .\end{equation} Replacing the above to $V_{boson}$
after some calculations we get:
\begin{align}\label{51}
V_{boson}&=-\frac{1}{2}\frac{\sqrt{\pi }}{(2\pi )^{d}a}(2\pi )^{\frac{d-1}{2}%
}m^{d+1}\Gamma (-\nu -\frac{1}{2}+1)\\&\notag -\frac{\sqrt{\pi }}{(2\pi )^{d}a}(2\pi )^{\frac{d-1}{2}%
}m^{d+1}\Gamma (-\nu -\frac{1}{2}+1)(a^{2})^{\frac{1}{2}-\nu
}\\&\notag \times[\sum_{k=1}^{\infty
}(a^{2}+4\pi^{2}k^{2})^{\nu +\frac{1}{2}-1}]\\&\notag +\frac{1}{4}\frac{1}{(2\pi )^{d}}(2\pi )^{\frac{d-1}{2}%
}m^{d+1}\Gamma (-\nu ) .\end{align} We use the binomial expansion
(in the case that $d$ is even) or the Taylor expansion (in the
case $d$ odd) \cite{qradstein7}:
\begin{equation}\label{52}
(a^{2}+b^{2})^{\nu -\frac{1}{2}}=\sum_{l=0}^{\sigma }\frac{(\nu -\frac{1}{2})!}{(\nu -%
\frac{1}{2}-l)!l!}(a^{2})^{l}(b^{2})^{\nu -\frac{1}{2}-l}
.\end{equation} If $d$ is even, then $\sigma$ equals to,
\begin{equation}\label{53}
\sigma =\nu -\frac{1}{2} .\end{equation} If $d$ is odd then
$\sigma$ $\epsilon$ $N^*$. We shall deal both cases. Replacing the
sum into $V_{boson}$,
\begin{align}\label{54}
V_{boson}&=-\frac{1}{2}\frac{\sqrt{\pi }}{(2\pi )^{d}a}(2\pi )^{\frac{d-1}{2}%
}m^{d+1}\Gamma (-\nu -\frac{1}{2}+1)\\&\notag+
\frac{1}{4}\frac{1}{(2\pi )^{d}}(2\pi
)^{\frac{d-1}{2}}m^{d+1}\Gamma (-\nu )\\& \notag
-\frac{\sqrt{\pi }}{(2\pi )^{d}a}(2\pi )^{\frac{d-1}{2}%
}m^{d+1}\Gamma (-\nu -\frac{1}{2}+1)(a^{2})^{\frac{1}{2}-\nu }\\&
\notag \times [\sum_{k=1}^{\infty
}\sum_{l=0}^{\sigma }\frac{((2\pi )^{2})^{\nu -\frac{1}{2}-l}(\nu -\frac{1}{2})!}{%
(\nu -\frac{1}{2}-l)!l!}(a^{2})^{l}(k^{2})^{\nu -\frac{1}{2}-l}]
.\end{align} The last expression shall be the initial point for
the following two subsections.

\noindent A much more elegant computation involves the analytic
continuation of the Epstein-zeta function
\cite{elizalde,elizaldenew1,elizalde2,kirstennew1,kirsten14,kirsten12,gongcharov,odintsovelizalde,kirstenepstein,elizaldekirstenarme}.
In a following section we shall present the Epstein zeta functions
in much more detail. In our case, relation (\ref{51}) can be
written in a much more elegant way, using the one dimensional
Epstein zeta function,
\begin{equation}\label{onedimensionalepsteinzeta12}
Z_1^{m^2}(\nu,w,\alpha)=\sum_{n=1}^{\infty}\Big{[}w(n+\alpha)^2+m^2\Big{]}^{-\nu}
,\end{equation} In our case, $\alpha=0$. Particularly one can make
the relevant substitutions in the sum,
\begin{equation}\label{kostantaras}
\sum_{k=1}^{\infty }(a^{2}+4\pi^{2}k^{2})^{\nu +\frac{1}{2}-1}],
\end{equation}
in terms of the one dimensional Epstein zeta function,
(\ref{onedimensionalepsteinzeta12}).

\subsubsection{The Chowla-Selberg Formula}

It worths mentioning at this point a very important formula
related with the Bessel sums \cite{elizalde,elizalde2,kirstennew1}
of relation,
\begin{align}\label{chowla}
V_{boson}&=-\sum\limits_{q=1}^{\infty }\frac{1}{(2\pi )^{d}}(2\pi
)^{\frac{d-1}{2}}m^{d+1}\frac{K_{\frac{d+1}{2}}(\frac{mq}{T})}{(\frac{mq}{2T})^{\frac{d+1}{2}}}\\&
\notag =-\frac{1}{2}\sum\limits_{q=-\infty }^{\infty \prime
}\frac{1}{(2\pi )^{d}}(2\pi
)^{\frac{d-1}{2}}m^{d+1}\frac{K_{\frac{d+1}{2}}(\frac{mq}{T})}{(\frac{mq}{2T})^{\frac{d+1}{2}}}.
\end{align}
Apart from the inhomogeneous Epstein zeta
\cite{elizalde,elizaldenew1,elizalde2,kirstennew1,kirsten14,kirsten12,gongcharov,odintsovelizalde,kirstenepstein,elizaldekirstenarme},
there exist in the literature a generalization of the
inhomogeneous Epstein zeta function, namely the extended
Chowla-Selberg formula \cite{elizalde}, which we briefly describe
at this point. We start with a two dimensional generalization of
the Epstein zeta function,
\begin{equation}\label{generalizedepsteinzeta}
E(s;a,b,c;q)=\sum_{n,{\,}m{\,}\epsilon Z}=(am^2+bmn+cn^2+q)^{-q}.
\end{equation}
In the following $Q$ is equal to,
\begin{equation}\label{qadratrik}
Q(m,n)=am^2+bmn+cn^2 ,\end{equation} and also $\Delta$ is,
\begin{equation}\label{Delta}
\Delta =4ac-b^2.
\end{equation}
Following \cite{elizalde}, relation
(\ref{generalizedepsteinzeta}), can be written as,
\begin{align}\label{choola}
&E(s;a,b,c;q)=2\zeta_{EH}(s,q/a)a^{-s}+\frac{2^{2s}\sqrt{\pi}a^{s-1}}{\Gamma
(s)\sqrt{a}}\sum_{n=1}^{\infty}\Gamma(s-1/2)\zeta_{EH}(s-1/2,4aq/\Delta)
\\ &\notag +\frac{2^{2s}\sqrt{\pi}a^{s-1}}{\Gamma
(s)\sqrt{a}}\sum_{n=1}^{\infty}n^{s-1/2}\cos(n\pi
b/a)\sum_{d/n}d^{1-2s}\Big{(}\Delta+\frac{4aq}{d^2}\Big{)}^{1/4-s/2}K_{s-1/2}\Big{(}\frac{\pi
n}{a}\sqrt{\Delta+\frac{4aq}{d^2}}\Big{)}.
\end{align}
In the above relation, the summation $\sum_{d/n}$ is over the
$1-2s$ powers of the divisors of $n$. Also $\zeta_{EH}$ stands
for,
\begin{align}\label{epsteinzhurwitzeta}
\zeta_{EH}(s;p)=-\frac{p^{-s}}{2}+\frac{\sqrt{\pi \Gamma
(s-1/2)}}{2\Gamma
(s)}p^{-s+1/2}+\frac{2\pi^{s}p^{-s/2+1/4}}{\Gamma
(s)}\sum_{n=1}^{\infty}n^{s-1/2}K_{s-1/2}(2\pi n\sqrt{p}).
\end{align}
Relation (\ref{choola}) has very attractive features. Most
importantly the exponential convergence. We just mention this here
for completeness and because (\ref{choola}) is very important. For
more details see the detailed description of \cite{elizalde}. Our
case is a special case of the extended Chowla-Selberg formula.

\subsubsection{The Case $d$ odd}

As stated before in the $d$ odd case, $\sigma $ $\epsilon$ $N^*$.
Then $V_{boson}$ is:
\begin{align}\label{55}
V_{boson}&=-\frac{1}{2}\frac{\sqrt{\pi }}{(2\pi )^{d}a}(2\pi )^{\frac{d-1}{2}%
}m^{d+1}\Gamma (-\nu -\frac{1}{2}+1)\\& \notag
+\frac{1}{4}\frac{1}{(2\pi )^{d}}(2\pi
)^{\frac{d-1}{2}}m^{d+1}\Gamma (-\nu )\\& \notag -\frac{\sqrt{\pi }}{(2\pi )^{d}a}(2\pi )^{\frac{d-1}{2}%
}m^{d+1}\Gamma (-\nu -\frac{1}{2}+1)(a^{2})^{\frac{1}{2}-\nu }\\&
\notag \times [\sum_{k=1}^{\infty
}\sum_{l=0}^{\sigma }\frac{((2\pi )^{2})^{\nu -\frac{1}{2}-l}(\nu -\frac{1}{2})!}{%
(\nu -\frac{1}{2}-l)!l!}(a^{2})^{l}(k^{2})^{\nu -\frac{1}{2}-l}].
\end{align}
Using the analytic continuation of the Riemann zeta function
\cite{elizalde,elizalde2,kirstennew1,tich,elizaldenew1},
\begin{equation}\label{56}
\zeta (s)=\sum_{n=1}^{\infty }n^{-s} ,\end{equation} to negative
integers, $V_{boson}$ becomes:
\begin{align}\label{57}
V_{boson}&=-\frac{1}{2}\frac{\sqrt{\pi }}{(2\pi )^{d}a}(2\pi )^{\frac{d-1}{2}%
}m^{d+1}\Gamma (-\nu -\frac{1}{2}+1)
\\& \notag +\frac{1}{4}\frac{1}{(2\pi )^{d}}(2\pi
)^{\frac{d-1}{2}}m^{d+1}\Gamma (-\nu )\\& \notag -\frac{\sqrt{\pi }}{(2\pi )^{d}a}(2\pi )^{\frac{d-1}{2}%
}m^{d+1}\Gamma (-\nu -\frac{1}{2}+1)(a^{2})^{\frac{1}{2}-\nu }\\& \notag \times[\sum_{l=0}^{\sigma }%
\frac{((2\pi )^{2})^{\nu -\frac{1}{2}-l}(\nu -\frac{1}{2})!}{(\nu -\frac{1}{2}-l)!l!}%
(a^{2})^{l}\zeta (-2\nu  +1+2l)] .\end{align} This is the final
form of the bosonic contribution to the effective potential for
$d$ odd. In the following we compute the above in the case $d=3$.
This will be done by Taylor expanding the last expression in
powers of $\varepsilon$ (with $d=3+\varepsilon$) as $\varepsilon
\rightarrow 0$.

\noindent Let us explicitly show how the poles are erased. In the
case $d=3$ two terms of $V_{boson}$ have poles. The first pole
appears in $\Gamma (-\nu )$ (remember $\ \nu =\frac{d+1}{2}$) and
the other is contained in $\zeta (-2\nu +1+2l)$ for the value
$l=2$ that gives the pole of $\zeta (s)$ for $s=1$. These terms
expanded around $d=3+\epsilon $, in the limit $\epsilon
\rightarrow 0$ are written:
\begin{eqnarray}\label{58}
\frac{1}{4}\frac{1}{(2\pi )^{d}}(2\pi )^{\frac{d-1}{2}}m^{d+1}\Gamma (-\nu ) &=&%
\frac{-m^{4}}{16\,\pi ^{2}\,\varepsilon }+(\frac{3\,m^{4}}{64\,\pi ^{2}}-\frac{\gamma \,m^{4}}{32\,\pi ^{2}}\\
&&+\frac{%
m^{4}\,\ln (2)}{32\,\pi ^{2}}-\frac{m^{4}\,\ln (m)}{16\,\pi ^{2}}\nonumber \\
&&+\frac{%
m^{4}\,\ln (\pi )}{32\,\pi ^{2}})+O(\varepsilon )\nonumber
\end{eqnarray}
(where $\gamma$ the Euler-Masceroni constant) in which a pole
appears,
\begin{equation}\label{59}
\frac{-m^{4}}{16\,\pi ^{2}\,\varepsilon } .\end{equation}
Regarding the other pole containing term (for $d=3+\epsilon$, $
\epsilon \rightarrow 0$),
\begin{align}\label{60}
&-\frac{\sqrt{\pi }}{(2\pi )^{d}a}(2\pi
)^{\frac{d-1}{2}}m^{d+1}\Gamma (-\nu
-\frac{1}{2}+1)(a^{2})^{\frac{1}{2}-\nu }\\&\notag
\times[\frac{((2\pi )^{2})^{\nu -\frac{1}{2}-2}(\nu
-\frac{1}{2})!}{(\nu -\frac{1}{2}-2)!l!} (a^{2})^{2}\zeta (-2\nu
+1+4)]=\\&\notag \frac{m^{4}\,}{16\,\pi ^{2}\,\,\varepsilon
}+\frac{-\left( \gamma \,m^{4}\,\right) }{16\,\pi
^{2}}+\frac{m^{4}\,\,\ln (2)}{32\,\pi ^{2}}\\&\notag
+\frac{m^{4}\,\,\ln (m)}{16\,\pi ^{2}}+\frac{m^{4}\,\,\ln (\pi
)}{32\,\pi ^{2}}-
\\&\notag \frac{m^{4}\,\,\ln (\alpha ^{2})}{32\,\pi
^{2}}-\frac{m^{4}\,\,\psi (\frac{1}{2})}{32\,\pi ^{2}}
\\&\notag +\frac{m^{4}\,\,\psi (\frac{5}{2})}{32\,\pi ^{2}
})+O(\varepsilon ) ,\end{align} with $\psi$ the digamma function.
Summing the above expressions we observe that the poles are
naturally erased as a consequence of the zeta regularization
method.

\noindent We expand $V_{boson}$ keeping the most dominant terms in
the high temperature limit
\cite{elizalde,kirstennew1,kapusta,Das}:
\begin{eqnarray}\label{61}
V_{boson} &=&\frac{\frac{-m^{4}}{16\,\pi ^{2}}+\frac{m^{4}\,\sqrt{\alpha ^{2}}}{%
16\,\pi ^{2}\,\alpha }}{\varepsilon }+(\frac{3\,m^{4}}{64\,\pi ^{2}}-\frac{%
\gamma \,m^{4}}{32\,\pi ^{2}}-\frac{m^{4}}{6\,\pi \,\alpha }-\frac{%
m^{4}\,\pi ^{2}}{45\,\alpha ^{4}}\nonumber \\
&&+\frac{m^{4}}{12\,\alpha ^{2}}-\frac{\gamma \,m^{4}\,}{16\,\pi
^{2}\,}+\frac{m^{4}\,\ln (2)}{16\,\pi
^{2}}+\frac{%
m^{4}\,\ln (\pi )}{16\,\pi ^{2}}\nonumber \\
&&-\frac{m^{4}\,\ln (\alpha ^{2})%
}{32\,\pi ^{2}\,}-\frac{m^{4}\,\psi (-%
\left( \frac{3}{2}\right) )}{32\,\pi ^{2}\,}\nonumber\\
&& -\frac{m^{4}\,\psi (\frac{1}{2})}{32\,\pi
^{2}\,}+\frac{m^{4}\,\psi (\frac{5}{2})}{%
32\,\pi ^{2}\,})+O(\varepsilon )
\end{eqnarray}
and substituting $\alpha =\frac{m}{T}$ we get:
\begin{eqnarray}\label{62}
V_{boson} &=&\frac{\frac{-m^{4}}{16\,\pi ^{2}}+\frac{m^{4}\,}{16\,\pi ^{2}}}{%
\varepsilon }+(\frac{3\,m^{4}}{64\,\pi ^{2}}-\frac{\gamma
\,m^{4}}{32\,\pi ^{2}}-\frac{m^{3}\,T}{6\,\pi }-\frac{\gamma
\,m^{4}\,}{16\,\pi ^{2}}+\nonumber
\\&&\frac{m^{2}\,\,T^{2}}{12}-\frac{\pi
^{2}\,\,T^{4}}{45\,}+\frac{m^{4}\,\ln (2)}{16\,\pi
^{2}}+\frac{%
m^{4}\,\ln (\pi )}{16\,\pi ^{2}}\nonumber \\
&&-\frac{m^{4}\,\ln (\alpha ^{2})%
}{32\,\pi ^{2}\,}-\frac{m^{4}\,\psi (-%
\left( \frac{3}{2}\right) )}{32\,\pi ^{2}\,}\nonumber\\
&& -\frac{m^{4}\,\psi (\frac{1}{2})}{32\,\pi
^{2}\,}+\frac{m^{4}\,\psi (\frac{5}{2})}{%
32\,\pi ^{2}\,})+O(\varepsilon ) .\end{eqnarray} In equation
(\ref{62}) we kept terms of order $\sim T$. For $\sigma =8$ we
have additionally,
\begin{align}\label{63}
&\frac{-\left( m^{7}\,\frac{m}{T}\,\zeta (5)\right) }{%
4096\,\pi ^{6}\,T^{3}}+\frac{m^{9}\,\frac{m}{T}\,\zeta (7)}{%
32768\,\pi ^{8}\,T^{5}}-\frac{7\,m^{11}\,\frac{m}{T}\,\zeta
(9)}{1572864\,\pi ^{10}\,T^{7}}
\\&\notag +\frac{3\,m^{13}\,\frac{m}{T}%
\,\zeta (11)}{4194304\,\pi
^{12}\,T^{9}}-\frac{33\,m^{15}\,\frac{m}{T}\,\zeta
(13)}{268435456\,\pi ^{14}\,T^{11}}) .\end{align}


\subsubsection{The Case $d$ even}
In the case $d$ even, $\sigma$ takes a limited number of values.
Particularly all the integer values up to the number $\sigma
=v-\frac{1}{2}$. Before proceeding we comment on the values that
$d$ can take. If it takes values $d>2$ that is $4,6..$, the theory
ceases to be renormalizable and UV regulators must be used in
order to cure UV singularities \cite{Das,kapusta,bordagreview}. We
shall not deal with these problems that usually appear in extra
dimensional models. Now $V_{boson}$ in the $d$ even case becomes:
\begin{align}\label{64}
V_{boson}&=-\frac{1}{2}\frac{\sqrt{\pi }}{(2\pi )^{d}a}(2\pi
)^{\frac{d-1}{2}}m^{d+1}\Gamma (-v-\frac{1}{2}+1)\\& \notag
+\frac{1}{4}\frac{1}{(2\pi )^{d}}(2\pi
)^{\frac{d-1}{2}}m^{d+1}\Gamma (-\nu )\\&\notag -\frac{\sqrt{\pi
}}{(2\pi )^{d}a}(2\pi )^{\frac{d-1}{2}}m^{d+1}\Gamma (-\nu
-\frac{1}{2}+1)(a^{2})^{\frac{1}{2}-\nu }\\&\notag \times
[\sum_{k=1}^{\infty }\sum_{l=0}^{\nu -\frac{1}{2}}\frac{((2\pi
)^{2})^{\nu -\frac{1}{2}-l}(\nu -\frac{1}{2})!}{(\nu
-\frac{1}{2}-l)!l!}(a^{2})^{l}(k^{2})^{\nu -\frac{1}{2}-l}]
,\end{align} and using the zeta regularization
\cite{elizalde,elizalde2,kirstennew1,elizaldenew1} we get:
\begin{align}\label{65}
V_{boson}&=-\frac{1}{2}\frac{\sqrt{\pi }}{(2\pi )^{d}a}(2\pi )^{\frac{d-1}{2}%
}m^{d+1}\Gamma (-\nu
-\frac{1}{2}+1)\\&\notag+\frac{1}{4}\frac{1}{(2\pi )^{d}}(2\pi
)^{\frac{d-1}{2}}m^{d+1}\Gamma (-\nu )\\&\notag -\frac{\sqrt{\pi }}{(2\pi )^{d}a}(2\pi )^{\frac{d-1}{2}%
}m^{d+1}\Gamma (-\nu -\frac{1}{2}+1)(a^{2})^{\frac{1}{2}-\nu }\\&\notag \times [\sum_{l=0}^{\nu -\frac{%
1}{2}}\frac{((2\pi )^{2})^{\nu -\frac{1}{2}-l}(\nu
-\frac{1}{2})!}{(\nu -\frac{1}{2} -l)!l!}(a^{2})^{l}\zeta (-2\nu
+1+2l)] .\end{align} We compute for example the above in the case
$d=2$. We can easily see that the poles are contained in the terms
$\Gamma (-\nu -\frac{1}{2}+1)$ and $\Gamma (-\nu -\frac{1}{2}+1)$.
Expanding for $\varepsilon \rightarrow 0 $ ($d=2+\varepsilon $)
the first pole containing term is:
\begin{align}\label{66}
-&\frac{1}{2}\frac{\sqrt{\pi }}{(2\pi )^{d}a}(2\pi )^{\frac{d-1}{2}%
}m^{d+1}\Gamma (-\nu -\frac{1}{2}+1)=\frac{-\left( m^{2}\,T\right) }{2\,%
\sqrt{2}\,\pi \,\varepsilon }+(\frac{m^{2}\,T}{4\,\sqrt{2}\,\pi }- \\
&\notag \frac{\gamma m^{2}\,T}{4\,\sqrt{2}\,\pi }+\frac{m^{2}\,T\,\ln (2)}{4\,%
\sqrt{2}\,\pi }-\frac{m^{2}\,T\,\ln (m)}{2\,\sqrt{2}\,\pi }+\frac{%
m^{2}\,T\,\ln (\pi )}{4\,\sqrt{2}\,\pi }) ,\end{align} and the
other one reads:
\begin{align}\label{67}
&-\frac{\sqrt{\pi }}{(2\pi )^{d}a}(2\pi )^{\frac{d-1}{2}%
}m^{d+1}\Gamma (-\nu -\frac{1}{2}+1)(a^{2})^{\frac{1}{2}-\nu }\\&\notag \times [\sum_{l=0}^{\nu -\frac{%
1}{2}}\frac{((2\pi )^{2})^{\nu -\frac{1}{2}-l}(\nu -\frac{1}{2})!}{(\nu -\frac{1}{2}%
-l)!l!}(a^{2})^{l}\zeta (-2\nu  +1+2l)]=\\&\notag \frac{m^{2}\,T}{2\,\sqrt{2}\,\pi \,\varepsilon }+(\frac{\gamma \,m^{2}\,T}{%
4\,\sqrt{2}\,\pi }+\frac{m^{2}\,T\,\ln (2)}{4\,\sqrt{2}\,\pi }+\frac{%
m^{2}\,T\,\ln (m)}{2\,\sqrt{2}\,\pi }\\&\notag
+\frac{m^{2}\,T\,\ln (\pi )}{4\,\sqrt{2}\,\pi
}-\frac{m^{2}\,T\,\ln
(2\,\pi )}{2\,\sqrt{2}\,\pi }-\frac{m^{2}\,T\,\ln (\frac{m^{2}}{T^{2}})}{4\,%
\sqrt{2}\,\pi }+2\,\sqrt{2}\,\pi \,T^{3}\,\zeta ^{\prime }(-2))
.\end{align} Adding equation (\ref{66}) and (\ref{67}) we can see
that the poles are erased naturally and $V_{boson}$ becomes
($d=2$):
\begin{eqnarray}\label{68}
V_{boson} &=&(\frac{m^{3}}{6\,\sqrt{2}\,\pi }+\frac{m^{2}\,T}{4\,\sqrt{2}\,\pi }%
+\frac{m^{2}\,T\,\ln (2)}{2\,\sqrt{2}\,\pi }+ \\
&&\frac{m^{2}\,T\,\ln (\pi )}{2\,\sqrt{2}\,\pi
}-\frac{m^{2}\,T\,\ln
(2\,\pi )}{2\,\sqrt{2}\,\pi }\nonumber \\
&& -\frac{m^{2}\,T\,\ln (\frac{m^{2}}{T^{2}})}{4\,%
\sqrt{2}\,\pi }+2\,\sqrt{2}\,\pi \,T^{3}\,\zeta ^{\prime
}(-2))\nonumber .\end{eqnarray}

\subsection{Fermionic Contribution at Finite Temperature}

In this section we will compute the fermionic contribution to the
effective potential:
\begin{equation}\label{69}
T\int \frac{dk^{3}}{(2\pi )^{3}}\sum_{n=-\infty }^{\infty }\ln
[(2n+1)^{2}\pi^{2}T^{2}+k^{2}+m^{2}].
\end{equation}
Following the same procedures as in the bosonic case we obtain
\cite{kirstennew1,elizalde,elizalde2,Das,kapusta}:
\begin{align}\label{70}
&T\int \frac{dk^{3}}{(2\pi )^{3}}\sum \ln [(2n+1)^{2}\pi^{2}T^{2}+k^{2}+m^{2}]=\\&\notag\int \frac{dk^{3}}{(2\pi )^{3}}\int\limits_{-%
\infty }^{\infty }\frac{dx}{2\pi }\ln [x^{2}+a^{2}]+2T\int
\frac{dk^{3}}{(2\pi )^{3}}\ln [1+e^{-(\frac{a}{2T})}].
\end{align}
As before, the first term to the left hand side is the effective
potential at zero temperature. We shall dwell on the temperature
dependent contribution, which in $d+1$ dimensions is written,
\begin{align}\label{71}
&T\int \frac{dk^{d}}{(2\pi )^{d}}\sum \ln
[4\pi^{2}n^{2}T^{2}+k^{2}+m^{2}]=\\& \notag \int
\frac{dk^{d+1}}{(2\pi )^{d+1}}\ln [k^{2}+a^{2}]+2T\int
\frac{dk^{d}}{(2\pi )^{d}}\ln [1+e^{-(\frac{a}{2T})}].
\end{align}
Let,
\begin{equation}\label{72}
V_{fermion}=2T\int \frac{dk^{d}}{(2\pi )^{d}}\ln
[1+e^{-(\frac{a}{2T})}]=2T\int \frac{dk^{d}}{(2\pi )^{d}}\ln
[1+e^{-\frac{a}{2T}}] .\end{equation} By using \cite{qradstein7},
\begin{equation}\label{}
\ln [1+e^{-\frac{a}{2T}}]=-\sum\limits_{q=1}^{\infty }\frac{(-1)^{q}e^{-%
\frac{a}{2T}q}}{q} ,\end{equation} $V_{fermion}$ becomes,
\begin{eqnarray}\label{73}
V_{fermion} &=&2T\int \frac{dk^{d}}{(2\pi )^{d}}\ln [1+e^{-\frac{a}{2T}%
}]\\&=&-2T\int \frac{dk^{d}}{(2\pi )^{d}}\sum\limits_{q=1}^{\infty }\frac{%
(-1)^{q}e^{-\frac{a}{2T}q}}{q} \nonumber \\
&=&-2\sum\limits_{q=1}^{\infty }T\int \frac{dk^{d}}{(2\pi )^{d}}\frac{%
(-1)^{q}e^{-\frac{a}{2T}q}}{q}  \nonumber .\end{eqnarray} Recall
that,
\begin{equation}\label{74}
a=\sqrt{k^{2}+m^{2}} ,\end{equation} and so,
\begin{eqnarray}\label{75}
V_{fermion} &=&-2\sum\limits_{q=1}^{\infty }T\int \frac{dk^{d}}{(2\pi )^{d}}%
\frac{(-1)^{q}e^{-\frac{\sqrt{k^{2}+m^{2}}}{2T}q}}{q}\nonumber \\
&=&-2\sum\limits_{q=1}^{%
\infty }T\int_{-\infty }^{\infty }\frac{dk}{(2\pi
)^{d}}k^{d-1}\frac{(2\pi
)^{\frac{d}{2}}}{\Gamma (\frac{d}{2})}\frac{(-1)^{q}e^{-\frac{\sqrt{%
k^{2}+m^{2}}}{2T}q}}{q} \\
&=&-2\sum\limits_{q=1}^{\infty }T\frac{(2\pi )^{\frac{d}{2}}(-1)^{q}}{%
\Gamma (\frac{d}{2})q(2\pi )^{d}}\int_{-\infty }^{\infty }dkk^{d-1}e^{-%
\frac{\sqrt{k^{2}+m^{2}}}{2T}q}  \nonumber .\end{eqnarray} The
integral,
\begin{equation}\label{arxidomania}
\int_{-\infty }^{\infty
}dkk^{d-1}e^{-\frac{\sqrt{k^{2}+m^{2}}}{2T}q} ,\end{equation}
 equals to \cite{qradstein7},
\begin{equation}\label{76}
\int_{-\infty }^{\infty }dkk^{d-1}e^{-\frac{\sqrt{k^{2}+m^{2}}}{2T}q}=2^{%
\frac{d}{2}-1}(\sqrt{\pi })^{-1}(\frac{q}{2T})^{\frac{1}{2}-\frac{d}{2}}m^{%
\frac{d+1}{2}}\Gamma (\frac{d}{2})K_{\frac{d+1}{2}}(\frac{mq}{2T})
.\end{equation} So $V_{fermion}$ reads,
\begin{eqnarray}\label{77}
V_{fermion} &=&-2\sum\limits_{q=1}^{\infty }\frac{2^{\frac{d}{2}-1}(-1)^{q}}{%
(2\pi )^{d}}(2\pi )^{\frac{d+1}{2}}m^{d+1}\frac{K_{\frac{d+1}{2}}(\frac{mq}{2T})}{1%
}(\frac{2T}{mq})^{\frac{d+1}{2}} \\
&=&-\sum\limits_{q=1}^{\infty }\frac{(-1)^{q}}{(2\pi )^{d}}(2\pi )^{\frac{%
d-1}{2}}m^{d+1}\frac{K_{\frac{d+1}{2}}(\frac{mq}{2T})}{(\frac{mq}{4T})^{\frac{d+1}{%
2}}}  \nonumber .\end{eqnarray} Using the relation
\cite{qradstein7,elizalde,elizalde2,kirstennew1}:
\begin{equation}\label{78}
\sum_{q=1}^{\infty }(-1)^{q}f(r)=2\sum_{q=1}^{\infty
}f(2r)-\sum_{q=1}^{\infty }f(r) ,\end{equation} we get,
\begin{equation}\label{79}
\sum\limits_{q=1}^{\infty }\frac{(-1)^{q}K_{\frac{d+1}{2}}(\frac{mq}{2T})}{(\frac{%
mq}{4T})^{\frac{d+1}{2}}}=2\sum\limits_{q=1}^{\infty }\frac{K_{\frac{d+1}{2}}(%
\frac{mq}{T})}{(\frac{mq}{2T})^{\frac{d+1}{2}}}-\sum\limits_{q=1}^{\infty }%
\frac{K_{\frac{d+1}{2}}(\frac{mq}{2T})}{(\frac{mq}{4T})^{\frac{d+1}{2}}}
.\end{equation} and upon replacing to $V_{fermion}$ we obtain:
\begin{eqnarray}\label{80}
V_{fermion}&=&-\sum\limits_{q=1}^{\infty }\frac{(-1)^{q}}{(2\pi )^{d}}(2\pi )^{%
\frac{d-1}{2}}m^{d+1}\frac{K_{\frac{d+1}{2}}(\frac{mq}{2T})}{(\frac{mq}{4T})^{%
\frac{d+1}{2}}} \\
&=&-\frac{(2\pi )^{\frac{d-1}{2}}m^{d+1}}{(2\pi )^{d}}{\Large (}%
2\sum\limits_{q=1}^{\infty }\frac{K_{\frac{d+1}{2}}(\frac{mq}{T})}{(\frac{mq}{2T}%
)^{\frac{d+1}{2}}}-\sum\limits_{q=1}^{\infty }\frac{K_{\frac{d+1}{2}}(\frac{mq}{2T%
})}{(\frac{mq}{4T})^{\frac{d+1}{2}}}{\Large )}  \nonumber
.\end{eqnarray} The function,
\begin{equation}\label{81}
\frac{K_{\nu }(z)}{(\frac{z}{2})^{\nu
}}=\frac{1}{2}\int_{0}^{\infty }
\frac{e^{-t-\frac{z^{2}}{4t}}}{t^{\nu +1}}dt] ,\end{equation} is
even under the transformation $z\rightarrow -z$. Thus the above
becomes:
\begin{align}\label{82}
&V_{fermion} =-\frac{(2\pi )^{\frac{d-1}{2}}m^{d+1}}{(2\pi )^{d}}{\Big{(}}2%
\frac{1}{2}\sum\limits_{q=-\infty }^{\infty \prime}\frac{K_{\frac{d+1}{2}}(\frac{mq}{T}%
)}{(\frac{mq}{2T})^{\frac{d+1}{2}}}\\&\notag
-\frac{1}{2}\sum\limits_{q=-\infty }^{\infty \prime}\frac{K_{\frac{d+1}{2}}(\frac{mq}{2T})}{(\frac{mq}{4T})^{\frac{d+1}{2}}%
}{\Big{)}} \\&\notag =-\frac{(2\pi )^{\frac{d-1}{2}}m^{d+1}}{(2\pi )^{d}}{\Large (}%
\sum\limits_{q=-\infty }^{\infty \prime}\frac{K_{\frac{d+1}{2}}(\frac{mq}{T})}{(\frac{%
mq}{2T})^{\frac{d+1}{2}}}-\frac{1}{2}\sum\limits_{q=-\infty }^{\infty \prime}%
\frac{K_{\frac{d+1}{2}}(\frac{mq}{2T})}{(\frac{mq}{4T})^{\frac{d+1}{2}}}{\Large)}
,\end{align} where the symbol $ ^{\prime}$ denotes omission of the
zero modes in the summation. Using,
\begin{equation}
\frac{K_{\nu }(z)}{(\frac{z}{2})^{\nu }}=\frac{1}{2}\int\limits_{0}^{\infty }%
\frac{e^{-t-\frac{z^{2}}{4t}}}{t^{\nu +1}}dt ,\end{equation} the
two Bessel sums are written:
\begin{align}\label{tade}
&-\frac{(2\pi )^{\frac{d-1}{2}}m^{d+1}}{(2\pi
)^{d}}\sum\limits_{q=-\infty }^{\infty \prime}\frac{K_{\frac{d+1}{2}}(\frac{mq}{T})}{(\frac{mq}{2T})^{\frac{d+1}{2}}}%
=\\&\notag -\frac{1}{2}\frac{1}{(2\pi )^{d}}(2\pi )^{\frac{d-1}{2}}m^{d+1}\int_{0}^{%
\infty }dte^{-t}\frac{\sum\limits_{q=-\infty }^{\infty \prime \prime }e^{-%
\frac{(\frac{mq}{T})^{2}}{4t}}}{t^{\frac{d+1}{2}+1}} .\end{align}
Set $\lambda =\frac{(\frac{m}{T})^{2}}{4t}$ and using the Poisson
summation formula \cite{elizalde,elizalde2,kirstennew1} we obtain:
\begin{equation}\label{83}
\ \sum_{q=-\infty }^{\infty \prime}e^{-\lambda q^{2}}=\sqrt{\frac{\pi }{\lambda }}%
(\ 1\ +\ \sum_{k=-\infty}^{\infty
\prime}e^{-\frac{4\pi^{2}k^{2}}{4\lambda }})\ -\ 1 .\end{equation}
Upon replacing we get:
\begin{align}\label{84}
&-\frac{(2\pi )^{\frac{d-1}{2}}m^{d+1}}{(2\pi
)^{d}}\sum\limits_{q=-\infty }^{\infty
\prime}\frac{K_{\frac{d+1}{2}}(\frac{mq}{T})}{(\frac{mq}{2T})^{\frac{d+1}{2}}}
=\\&\notag -\frac{1}{2}\frac{1}{(2\pi )^{d}}(2\pi )^{\frac{d-1}{2}}m^{d+1}\int_{0}^{%
\infty }dte^{-t}\Big{(}\frac{\sqrt{\frac{\pi }{\lambda }}\big{(}\
1\ +\ \sum_{k=-\infty}^{\infty
\prime}e^{-\frac{4\pi^{2}k^{2}}{4\lambda }}\big{)}\ -\
1}{t^{\frac{d+1}{2}+1}}\Big{)} .\end{align} Set,
\begin{equation}\label{85}
\nu =\frac{d+1}{2} .\end{equation} and thus,
\begin{align}\label{86}
&-\frac{(2\pi )^{\frac{d-1}{2}}m^{d+1}}{(2\pi
)^{d}}\sum\limits_{q=-\infty }^{\infty
\prime}\frac{K_{\frac{d+1}{2}}(\frac{mq}{T})}{(\frac{mq}{2T})^{\frac{d+1}{2}}}
=\\&\notag-\frac{1}{2}\frac{1}{(2\pi )^{d}}(2\pi )^{\frac{d-1}{2}%
}m^{d+1}(\int_{0}^{\infty }dte^{-t}\frac{\sqrt{\frac{\pi }{\lambda }}\ }{%
t^{\nu +1}}) \\&\notag -\frac{1}{2}\frac{1}{(2\pi )^{d}}(2\pi )^{\frac{d-1}{2}}m^{d+1}\int_{0}^{%
\infty }dte^{-t}(\frac{\sqrt{\frac{\pi }{\lambda }}(\
\sum_{k=-\infty}^{\infty \prime}e^{-\frac{4\pi^{2}k^{2}}{4\lambda }})\ }{t^{\nu +1}}) \\
&\notag+\frac{1}{2}\frac{1}{(2\pi )^{d}}(2\pi )^{\frac{d-1}{2}}m^{d+1}\int_{0}^{%
\infty }dte^{-t}(\frac{1}{t^{\nu +1}}) .\end{align} Also,
\begin{equation}\label{87}
a=\frac{m}{T} ,\end{equation} and finally (with $\lambda
=\frac{a^{2}}{4t}$),
\begin{align}\label{88}
&-\frac{(2\pi )^{\frac{d-1}{2}}m^{d+1}}{(2\pi
)^{d}}\sum\limits_{q=-\infty }^{\infty
\prime}\frac{K_{\frac{d+1}{2}}(\frac{mq}{T})}{(\frac{mq}{2T})^{\frac{d+1}{2}}}
=\\& \notag -\frac{\sqrt{\pi }}{(2\pi )^{d}a}(2\pi )^{\frac{d-1}{2}%
}m^{d+1}(\int_{0}^{\infty }dte^{-t}\ t^{-\nu
-\frac{1}{2}})\\&\notag
-\frac{\sqrt{\pi }}{(2\pi )^{d}a}(2\pi )^{\frac{d-1}{2}%
}m^{d+1}\int_{0}^{\infty }dte^{-t}(\frac{\sqrt{\pi t}2(\
\sum_{k=-\infty}^{\infty
\prime}e^{-\frac{4\pi^{2}k^{2}}{a^{2}}t})\ }{at^{\nu
+\frac{1}{2}}}) \\& \notag +\frac{1}{2}\frac{1}{(2\pi )^{d}}(2\pi
)^{\frac{d-1}{2}}m^{d+1}\int_{0}^{\infty }dte^{-t}(\frac{1}{t^{\nu
+1}}) .\end{align} By using \cite{qradstein7},
\begin{equation}\label{89}
\ \frac{1}{(x^{2}+a^{2})^{\mu +1}}=\frac{1}{\Gamma (\mu
+1)}\int_{0}^{\infty }dte^{-(x^{2}+a^{2})t}t^{\mu }
,\end{equation} we obtain the equation:
\begin{align}\label{90}
&-\frac{(2\pi )^{\frac{d-1}{2}}m^{d+1}}{(2\pi
)^{d}}\sum\limits_{q=-\infty }^{\infty
\prime}\frac{K_{\frac{d+1}{2}}(\frac{mq}{T})}{(\frac{mq}{2T})^{\frac{d+1}{2}}}
=-\frac{\sqrt{\pi }}{(2\pi )^{d}a}(2\pi
)^{\frac{d-1}{2}}m^{d+1}\Gamma (-\nu -\frac{1}{2}+1)\\&\notag
-\frac{\sqrt{\pi }}{(2\pi )^{d}a}(2\pi )^{\frac{d-1}{2}%
}m^{d+1}\Gamma (-\nu -\frac{1}{2}+1)[\sum_{k=-\infty}^{\infty
\prime}(1+(\frac{2\pi k}{a})^{2})^{\nu +\frac{1}{2}-1}]\\&\notag
+\frac{1}{2}\frac{1}{(2\pi )^{d}}(2\pi
)^{\frac{d-1}{2}}m^{d+1}\Gamma (-\nu ) .\end{align} The sum,
\begin{equation}\label{91}
\sum_{k=-\infty}^{\infty \prime}(1+(\frac{2\pi k}{a})^{2})^{\nu
+\frac{1}{2}-1} ,\end{equation} is invariant under the
transformation $k\rightarrow -k$. Thus we change the sum to,
\begin{equation}\label{92}
2\sum_{k=1}^{\infty }(1+(\frac{2\pi k}{a})^{2})^{\nu
+\frac{1}{2}-1} .\end{equation} Replacing again we get:
\begin{align}\label{93}
&-\frac{(2\pi )^{\frac{d-1}{2}}m^{d+1}}{(2\pi
)^{d}}\sum\limits_{q=-\infty }^{\infty
\prime}\frac{K_{\frac{d+1}{2}}(\frac{mq}{T})}{(\frac{mq}{2T})^{\frac{d+1}{2}}}=\\&\notag
-\frac{\sqrt{\pi }}{(2\pi )^{d}a}(2\pi
)^{\frac{d-1}{2}}m^{d+1}\Gamma (-\nu -\frac{1}{2}+1)\\&\notag
-\frac{2\sqrt{\pi }}{(2\pi )^{d}a}(2\pi
)^{\frac{d-1}{2}}m^{d+1}\Gamma (-\nu
-\frac{1}{2}+1)(a^{2})^{\frac{1}{2}-\nu }[\sum_{k=1}^{\infty
}(a^{2}+4\pi^{2}k^{2})^{\nu +\frac{1}{2}-1}]\\&\notag
\frac{1}{2}\frac{1}{(2\pi )^{d}}(2\pi
)^{\frac{d-1}{2}}m^{d+1}\Gamma (-\nu ) .\end{align} Using the
binomial expansion (in the case $d$ even) or Taylor expansion (in
the case $d$ odd) \cite{qradstein7}:
\begin{equation}\label{94}
(a^{2}+b^{2})^{\nu -\frac{1}{2}}=\sum_{l=0}^{\sigma }\frac{(\nu -\frac{1}{2})!}{(\nu -%
\frac{1}{2}-l)!l!}(a^{2})^{l}(b^{2})^{\nu -\frac{1}{2}-l}
.\end{equation} For $d$ even, $\sigma $ equals,
\begin{equation}\label{peos}
\sigma =\nu -\frac{1}{2} ,\end{equation} If $d$ is odd then
$\sigma$ is a positive integer. By Taylor expanding we obtain:
\begin{align}\label{95}
&-\frac{(2\pi )^{\frac{d-1}{2}}m^{d+1}}{(2\pi
)^{d}}\sum\limits_{q=-\infty }^{\infty \prime}\frac{K_{\frac{d+1}{2}}(\frac{mq}{T})}{(\frac{mq}{2T})^{\frac{d+1}{2}}}%
=\\&\notag -\frac{\sqrt{\pi }}{(2\pi )^{d}a}(2\pi )^{\frac{d-1}{2}%
}m^{d+1}\Gamma (-\nu -\frac{1}{2}+1)+\ \ \frac{1}{2}\frac{1}{(2\pi
)^{d}}(2\pi )^{\frac{d-1}{2}}m^{d+1}\Gamma (-\nu )\\&\notag \notag
-\frac{2\sqrt{\pi }}{(2\pi )^{d}a}(2\pi )^{\frac{d-1}{2}%
}m^{d+1}\Gamma (-\nu -\frac{1}{2}+1)(a^{2})^{\frac{1}{2}-\nu
}\\&\notag \times [\sum_{k=1}^{\infty
}\sum_{l=0}^{\sigma }\frac{((2\pi )^{2})^{\nu -\frac{1}{2}-l}(\nu -\frac{1}{2})!}{%
(\nu -\frac{1}{2}-l)!l!}(a^{2})^{l}(k^{2})^{\nu -\frac{1}{2}-l}]
.\end{align} Following the previous techniques we get for the
second sum of equation (\ref{tade}):
\begin{align}\label{96}
&\frac{1}{2}\frac{(2\pi )^{\frac{d-1}{2}}m^{d+1}}{(2\pi )^{d}}%
\sum\limits_{q=-\infty }^{\infty \prime}\frac{K_{\frac{d+1}{2}}(\frac{mq}{2T})}{(\frac{%
mq}{4T})^{\frac{d+1}{2}}}=\\&\notag
+\frac{1}{2}\frac{\sqrt{\pi}}{(2\pi )^{d}a_{1}}(2\pi
)^{\frac{d-1}{2}}m^{d+1}\Gamma (-\nu -\frac{1}{2}+1)
\\&\notag -\frac{1}{4}\frac{1}{(2\pi )^{d}}(2\pi
)^{\frac{d-1}{2}}m^{d+1}\Gamma (-\nu )\\&\notag
+\frac{1}{2}\frac{\sqrt{\pi }}{(2\pi )^{d}a_{1}}(2\pi )^{\frac{d-1}{2}%
}m^{d+1}\Gamma (-\nu -\frac{1}{2}+1)(a_{1}^{2})^{\frac{1}{2}-\nu }\\&\notag [\sum_{k=1}^{%
\infty }\sum_{l=0}^{\sigma }\frac{((2\pi )^{2})^{\nu -\frac{1}{2}-l}(\nu -\frac{1}{%
2})!}{(\nu -\frac{1}{2}-l)!l!}(a_{1}^{2})^{l}(k^{2})^{\nu
-\frac{1}{2}-l}] ,\end{align} with,
\begin{equation}\label{97}
 \alpha _{1}=\frac{m}{2T}
.\end{equation} Finally adding the resulting expressions we get:
\begin{align}\label{98}
&V_{fermion} =-\frac{(2\pi )^{\frac{d-1}{2}}m^{d+1}}{(2\pi )^{d}}{\Large (}%
\sum\limits_{q=-\infty }^{\infty \prime}\frac{K_{\frac{d+1}{2}}(\frac{mq}{T})}{(\frac{%
mq}{2T})^{\frac{d+1}{2}}}-\frac{1}{2}\sum\limits_{q=-\infty }^{\infty \prime}%
\frac{K_{\frac{d+1}{2}}(\frac{mq}{2T})}{(\frac{mq}{4T})^{\frac{d+1}{2}}}{\Large
)}
\notag \\&\notag =-\frac{\sqrt{\pi }}{(2\pi )^{d}a}(2\pi )^{\frac{d-1}{2}%
}m^{d+1}\Gamma (-\nu -\frac{1}{2}+1)+\ \ \frac{1}{2}\frac{1}{(2\pi
)^{d}}(2\pi )^{\frac{d-1}{2}}m^{d+1}\Gamma (-\nu )- \\&\notag
=-\frac{2\sqrt{\pi }}{(2\pi )^{d}a}(2\pi )^{\frac{d-1}{2}
}m^{d+1}\Gamma (-\nu -\frac{1}{2}+1)(a^{2})^{\frac{1}{2}-\nu
}\\&\notag \times [\sum_{k=1}^{\infty
}\sum_{l=0}^{\sigma }\frac{((2\pi )^{2})^{\nu -\frac{1}{2}-l}(\nu -\frac{1}{2})!}{%
(\nu -\frac{1}{2}-l)!l!}(a^{2})^{l}(k^{2})^{\nu
-\frac{1}{2}-l}]\\&\notag +\frac{1}{2}\frac{\sqrt{\pi }}{(2\pi
)^{d}a_{1}}(2\pi )^{\frac{d-1}{2} }m^{d+1}\Gamma (-\nu
-\frac{1}{2}+1)-\ \ \frac{1}{4}\frac{1}{(2\pi )^{d}}(2\pi
)^{\frac{d-1}{2}}m^{d+1}\Gamma (-\nu )\\&\notag
+\frac{1}{2}\frac{\sqrt{\pi }}{(2\pi )^{d}a_{1}}(2\pi
)^{\frac{d-1}{2} }m^{d+1}\Gamma (-\nu
-\frac{1}{2}+1)(a_{1}^{2})^{\frac{1}{2}-\nu }\\& \times
[\sum_{k=1}^{
\infty }\sum_{l=0}^{\sigma }\frac{((2\pi )^{2})^{\nu -\frac{1}{2}-l}(\nu -\frac{1}{%
2})!}{(\nu -\frac{1}{2}-l)!l!}(a_{1}^{2})^{l}(k^{2})^{\nu
-\frac{1}{2}-l}] ,\end{align} with $\alpha =\frac{m}{T}$ and
$\alpha _{1}=\frac{m}{2T}$. Using the zeta regularization
technique \cite{elizalde,elizalde2,elizaldenew1,kirstennew1,odi7}
we obtain,
\begin{align}\label{99}
&V_{fermion} =-\frac{(2\pi )^{\frac{d-1}{2}}m^{d+1}}{(2\pi )^{d}}{\Large (}%
\sum\limits_{q=-\infty }^{\infty \prime}\frac{K_{\frac{d+1}{2}}(\frac{mq}{T})}{(\frac{%
mq}{2T})^{\frac{d+1}{2}}}-\frac{1}{2}\sum\limits_{q=-\infty }^{\infty \prime}%
\frac{K_{\frac{d+1}{2}}(\frac{mq}{2T})}{(\frac{mq}{4T})^{\frac{d+1}{2}}}{\Large
)}
\\&\notag =-\frac{\sqrt{\pi }}{(2\pi )^{d}a}(2\pi )^{\frac{d-1}{2}%
}m^{d+1}\Gamma (-\nu -\frac{1}{2}+1)+\ \ \frac{1}{2}\frac{1}{(2\pi
)^{d}}(2\pi )^{\frac{d-1}{2}}m^{d+1}\Gamma (-\nu )\nonumber
\\&\notag -\frac{2\sqrt{\pi }}{(2\pi )^{d}a}(2\pi
)^{\frac{d-1}{2} }m^{d+1}\Gamma (-\nu
-\frac{1}{2}+1)(a^{2})^{\frac{1}{2}-\nu }\\&\notag \times
[\sum_{l=0}^{\sigma }
\frac{((2\pi )^{2})^{\nu -\frac{1}{2}-l}(\nu -\frac{1}{2})!}{(\nu -\frac{1}{2}-l)!l!}%
(a^{2})^{l}\zeta (-2\nu  +1+2l)]  \\&\notag
+\frac{1}{2}\frac{\sqrt{\pi }}{(2\pi )^{d}a_{1}}(2\pi
)^{\frac{d-1}{2} }m^{d+1}\Gamma (-\nu -\frac{1}{2}+1)-\ \
\frac{1}{4}\frac{1}{(2\pi )^{d}}(2\pi
)^{\frac{d-1}{2}}m^{d+1}\Gamma (-\nu ) \\&\notag
+\frac{1}{2}\frac{\sqrt{\pi }}{(2\pi )^{d}a_{1}}(2\pi
)^{\frac{d-1}{2} }m^{d+1}\Gamma (-\nu
-\frac{1}{2}+1)(a_{1}^{2})^{\frac{1}{2}-\nu }\\&\notag \times
[\sum_{l=0}^{
\sigma }\frac{((2\pi )^{2})^{\nu -\frac{1}{2}-l}(\nu -\frac{1}{2})!}{(\nu -\frac{1}{2}%
-l)!l!}(a_{1}^{2})^{l}\zeta (-2\nu  +1+2l)] .\end{align}

\noindent We kept the above expression without simplifying in
order to have a clear picture of the terms appearing (compare with
the bosonic case). In the case $d=3$ appear the poles we discussed
in the bosonic case. Again we Taylor expand around $d=3+\epsilon $
for $\epsilon \rightarrow 0$.

\noindent As in the bosonic case, we can write the fermionic
contribution at finite temperature more elegantly using the
analytic continuation of the Epstein-zeta
\cite{elizalde,elizaldenew1,elizalde2,kirstennew1,kirsten14,kirsten12,gongcharov,odintsovelizalde,kirstenepstein,elizaldekirstenarme}
function. In this case the sums of the form,
\begin{equation}\label{kostantaras}
\sum_{k=1}^{\infty }(a^{2}+4\pi^{2}(2k+1)^{2})^{\nu
+\frac{1}{2}-1}],\end{equation} can be written in terms of the one
dimensional Epstein zeta function,
\begin{equation}\label{onedimensionalepsteinzeta}
Z_1^{m^2}(\nu,w,\alpha)=\sum_{n=1}^{\infty}\Big{[}w(n+\alpha)^2+m^2\Big{]}^{-\nu}
,\end{equation} with $\alpha=\frac{1}{2}$ and so on. We postpone
the detailed presentation of the Epstein zeta functions in the
section in which we study the twisted boundary conditions
effective potential.

\subsubsection{Case $d$ odd}

For the case $d=3$, keeping terms $\sim T$ we have:
\begin{align}\label{100}
&V_{fermion}=\frac{\frac{-m^{4}}{16\,\pi ^{2}}+\frac{m^{4}\,}{16\,\pi ^{2}}}{%
\varepsilon }+\Big{(}\frac{3\,m^{4}}{64\,\pi ^{2}}-\frac{3\,\gamma
\,m^{4}}{32\,\pi
^{2}}\\&\notag -\frac{m^{2}\,T^{2}}{6}+\frac{%
14\,\pi ^{2}\,\,T^{4}}{45\,}+\frac{m^{4}\,\ln (\pi )}{16\,\pi
^{2}}\\&\notag -\frac{m^{4}\,\ln (\frac{m^{2}}{T^{2}})}{32\,\pi ^{2}}-\frac{%
m^{4}\,\,\psi (-\left( \frac{3}{2}\right) )}{32\,\pi ^{2}}-\frac{%
m^{4}\,\,\psi (\frac{1}{2})}{32\,\pi ^{2}}\\
&\notag +\frac{m^{4}\,\,\psi (\frac{5}{2})}{32\,\pi ^{2}}+\frac{%
7\,m^{6}\,\,\zeta (3)}{1536\,\pi ^{4}\,T^{2}}-\frac{31\,m^{8}\,\,\zeta (5)}{%
65536\,\pi ^{6}\,T^{4}}\Big{)} .\end{align} There are terms which
are inverse powers of the temperature which in the high
temperature limit (which we use) are negligible.

\bigskip

\subsubsection{Case $d$ Even}

The calculation is the same as in the bosonic case. We only quote
the case $d=2$

\begin{equation}\label{ypermalakas}
V_{fermion}=(\frac{m^{3}}{6\,\sqrt{2}\,\pi }-\frac{m^{2}\,T\,\ln (2)}{\sqrt{2}%
\,\pi }-12\sqrt{2}\,\pi \,T^{3}\,\zeta ^{\prime }(-2))
,\end{equation} We observe that the results contain a finite
number of terms and is not an infinite sum as in the case $d$ odd.

\subsection{Some Applications on Finite Temperature Field Theories}

\subsubsection{The Standard Model at Finite Temperature}

Let us now present the 1-loop correction for the effective
potential of standard model fields \cite{quirowfinite}. The
calculations of the final results are based on relations
(\ref{99}) and (\ref{54}), of the previous sections. We start with
a scalar boson described by the Lagrangian,
\begin{equation}\label{22revision}
    L=\frac{1}{2}\partial^\mu \phi \partial_\mu \phi-V_0(\phi)
,\end{equation} with tree level potential,
\begin{equation}\label{23revision}
    V_0=\frac{1}{2}m^2\phi^2+\frac{\lambda}{4!}\phi^4
,\end{equation} or in the case of $N_s$ complex scalar fields,
\begin{equation}\label{29revision}
    L=\frac{1}{2}\partial^\mu \phi^\alpha \partial_\mu \phi^\dag_\alpha-V_0(\phi^\alpha,\phi_\alpha^\dag)
,\end{equation} and in the following,
\begin{equation}\label{31revision}
    (M^2_s)_b^\alpha\equiv V_b^\alpha=\frac{\partial^2V}{\partial \phi_\alpha^\dag \partial \phi^b}
.\end{equation} Mention that $TrM_s^2=2V_\alpha^\alpha$, where $2$
comes from the two degrees of freedom that every complex scalar
field has. Also $TrI=2N_s$. Now regarding the fermion fields we
have,
\begin{equation}\label{32revision}
    L=i\overline{\psi}_\alpha \gamma \cdot \partial
    {\psi}^\alpha-\overline{\psi}_\alpha(M_f)^\alpha_b \psi^b
,\end{equation} where the mass matrix $(M_f)^\alpha_b(\phi_c^i)$,
is a function of scalar fields linear in $\phi_c^i$:
\begin{equation}\label{32arevision}
(M_f)^\alpha_b=\Gamma_{bi}^{\alpha}\phi_c^i.\end{equation} It is
assumed that a Higgs mechanism gives mass to fermions. Finally
consider the $SU(N)$ gauge invariant Lagrangian,
\begin{equation}\label{36revision}
    L=-\frac{1}{4}Tr(F_{\mu \nu}F^{\mu
    \nu})+\frac{1}{2}Tr(D_{\mu}\phi_{\alpha})^\dag(D_{\mu}\phi_{\alpha})....
,\end{equation} describing the gauge bosons-Higgs interactions. In
the following,

\begin{equation}\label{42arevision}
(M_{gb})^2_{\alpha
\beta}(\phi_c)=g_{\alpha}g_{\beta}Tr\Big{[}(T^i_{\alpha
l}\phi_i)^\dag T^l_{\beta j}\phi_j\Big{]} ,\end{equation} are the
gauge bosons masses, and $T_{\alpha}$ are the $SU(N)$ generators
in the adjoint representation. For the case of scalar bosons the
1-loop correction to the effective potential is,
\begin{equation}\label{159revisionistes}
V_{eff}^{\beta}(\phi_c)=V_0(\phi_c)+V_1^{\beta}(\phi_c)
,\end{equation} with $V_0(\phi_c)$ the tree order effective
potential and the loop correction,
\begin{equation}\label{160revisionistes}
V_1^{\beta}(\phi_c)=\frac{1}{2\beta}\sum_{n=-\infty}^{\infty}\int
\frac{\mathrm{d}^3p}{(2\pi)^3}\ln
{\Big{[}}\omega_n^2+\omega^2(\phi_c){\Big{]}} ,\end{equation}
where:
\begin{equation}\label{160arevisioninstes}
\omega_n=2n\pi \beta^{-1} ,\end{equation} and also
\begin{equation}\label{161revisionistes}
\omega^2=p^2+m^2(\phi_c).\end{equation} In the above,
$m^2(\phi_c)$ is given in relation (\ref{31revision}). Relation
(\ref{160revisionistes}) was the starting point of the our
calculation for the boson case, see relation (\ref{7}). Now in the
fermion case,
\begin{equation}\label{159brevisioninstes}
V_{eff}^{\beta}(\phi_c)=V_0(\phi_c)+V_1^{\beta}(\phi_c)
.\end{equation} where as before $V_0(\phi_c)$ the tree level
potential and $V_1^{\beta}(\phi_c)$ the 1-loop correction. The
last equals to
\begin{equation}\label{160frevisioninstes}
V_1^{\beta}(\phi_c)=-\frac{2\lambda}{2\beta}\sum_{n=-\infty}^{\infty}\int
\frac{\mathrm{d}^3p}{(2\pi)^3}\ln
{\Big{[}}\omega_n^2+\omega^2(\phi_c){\Big{]}} ,\end{equation} with
$\omega_n$ the fermionic Matsubara frequencies:
\begin{equation}\label{160hrevisioninstes}
\omega_n=(2n+1)\pi \beta^{-1} .\end{equation} Also,
\begin{equation}\label{161revisioninstes}
\omega^2=p^2+M_f^2(\phi_c) .\end{equation} Relation
(\ref{160frevisioninstes}) was the starting point for the fermion
effective potential calculation, relation (\ref{69}). Finally for
the gauge bosons case the tree effective potential with the 1-loop
correction reads,
\begin{equation}\label{211revisioninstes}
V_1^{\beta}(\phi_c)=Tr\Delta\Big{(}\frac{1}{2}\int
\frac{\mathrm{d}^4p}{(2\pi)^4}\ln
{\Big{[}}p^2+M_{gb}^2(\phi_c){\Big{]}}+\frac{1}{2\pi^2\beta^4}J_B[M_{gb}^2(\phi_c)\beta^2]\Big{)}
,\end{equation} where $Tr\Delta=3$. Notice that:
\begin{equation}\label{211arevisioninstes}
J_B[m^2\beta^2]=\int_0^{\infty} dx x^2
\ln[1-e^{-\sqrt{x^2+\beta^2m^2}}] ,\end{equation} and as before:
\begin{equation}\label{211brevisioninstes}
(M_{gb})^2_{\alpha
\beta}(\phi_c)=g_{\alpha}g_{\beta}Tr\Big{[}(T^i_{\alpha
l}\phi_i)^\dag T^l_{\beta j}\phi_j\Big{]} .\end{equation} Relation
(\ref{211arevisioninstes}) was obtained from relation (\ref{7}).

\subsection{Supersymmetric Effective Potential at Finite Temperature}

It is very useful to extend our analysis for scalar bosons,
fermions and gauge bosons in the supersymmetric case. Consider an
$N=1$, $d=4$ supersymmetric Lagrangian with an $SU(N)$ gauge
symmetry. After that we give a general formula for the
supersymmetric potential at finite temperature. We shall use the
$\overline{DR}^{\prime }$ renormalization scheme \cite{martin}.
The chiral superfield in components reads,
\begin{align}\label{62golftsi}
\Phi(x,\theta,\bar{\theta})&=A(x)+i\theta\sigma^\mu
\bar{\theta}\partial_\mu
A-\frac{1}{4}\theta^2\bar{\theta}^2\square A\\\notag
&+\sqrt{2}\theta \psi(x)-\frac{i}{\sqrt{2}}\theta \theta
\partial_\mu \psi \sigma^\mu \bar{\theta}+\theta \theta F(x)
,\end{align} and the vector hypermultiplet is described by the
chiral superfield,
\begin{equation}\label{65golftsi}
W_a=T^\alpha
\big{(}-\lambda^\alpha_a+\theta_aD^\alpha-\frac{i}{2}(\sigma^\mu\bar{\sigma}^\nu\theta)_aF^\alpha_{\mu
\nu}+\theta^2\sigma^\mu D_\mu \bar{\lambda}^\alpha \big{)}
,\end{equation} with,
\begin{equation}\label{65agolftsi}
F_{\mu \nu}^\alpha=\partial_\mu A_\nu^\alpha-\partial_\nu
A_\mu^\alpha+f^{\alpha bc}A_\mu^bA_\nu^c ,\end{equation} and also,
\begin{equation}\label{65agolftsi}
D_\mu\bar{\lambda}^\alpha=\partial_\mu \bar{\lambda}+f^{\alpha
bc}A_\mu^b\bar{\lambda}^c ,\end{equation} The $N=1$ Lagrangian is,
\begin{equation}\label{68golftsi}
L=\frac{1}{8\pi}Im\big{(}\tau Tr\int d^2 \theta W^\alpha
W_\alpha\big{)}+\int d^2\theta d^2 \bar{\theta}\Phi^\dag
e^{-2V}\Phi+\int d^2\theta W+\int d^2 \bar{\theta}
\bar{W}.\end{equation} which in components is written,
\begin{align}\label{69golftsi}
L&=-\frac{1}{4g^2}F_{\mu \nu}^\alpha F^{\alpha \mu
\nu}+\frac{\theta}{32\pi^2}F_{\mu \nu}^\alpha
\widetilde{F}^{\alpha \mu \nu}-\\\notag
&\frac{i}{g^2}\lambda^\alpha \sigma^\mu D_\mu
\bar{\lambda}^\alpha+\frac{1}{2g^2}D^\alpha D^\alpha+(\partial_\mu
A-iA_\mu^\alpha T^\alpha A)^\dag (\partial_\mu A-iA_\mu^\alpha
T^\alpha A)\\ \notag & -i\bar{\psi} \bar{\sigma}^\mu(\partial_\mu
\psi-iA_\mu^\alpha T^\alpha \psi)-D^\alpha A^\dag T^\alpha
A-i\sqrt{2}A^\dag T^\alpha \lambda^\alpha \psi\\ \notag &
+i\sqrt{2}\bar{\psi}T^\alpha A \bar{\lambda}^\alpha+F_i^\dag
F_i+\frac{\partial W}{\partial A_i}F_i+\frac{\partial
\bar{W}}{\partial A_i^\dag}F_i^\dag-\frac{1}{2}\frac{\partial
W}{\partial A_i \partial A_j}\psi_i
\psi_j-\frac{1}{2}\frac{\partial \bar{W}}{\partial A_i^\dag
\partial A_j^\dag}\bar{\psi}_i \bar{\psi}_j.
\end{align}
The computation of the finite temperature effective potential can
be done easily. The general potential up to one loop at finite
temperature is \cite{martin},
\begin{equation}\label{potentialmine1golftsi}
V=V_0+\frac{1}{64\pi^2}(V_{T=0}+V_{T\neq 0}) ,\end{equation} In
the above, $V_0$ is the tree order potential (appearing in the
Lagrangian). Also $V_{T=0}$ is the one loop effective potential at
$T=0$. It is given by:
\begin{align}\label{potentialmine2golftsi}
V_{T=0}&=\sum_{i}\Big{(}\ln(\frac{m_i^2}{Q^2}-\frac{3}{2})\Big{)}\\&
\notag+3\sum_j\Big{(}\ln(\frac{M_j^2}{Q^2}-\frac{3}{2})\Big{)}
-2\sum_k\Big{(}\ln(\frac{\mathcal{M}_{k}^2}{Q^2}-\frac{3}{2})\Big{)}.
\end{align}
Finally, $V_{T\neq 0}$, is given by:
\begin{align}\label{potentialmine3golftsi}
V_{T\neq 0}&=\sum_i\int
\frac{d^3k}{(2\pi)^3}2T\ln\big{(}1-e^{\frac{-\sqrt{k^2+m_i^2}}{T}}\big{)}\\
\notag &+3\sum_j\int
\frac{d^3k}{(2\pi)^3}2T\ln\big{(}1-e^{\frac{\sqrt{k^2+M_j^2}}{T}}\big{)}\\&
\notag -2\sum_k\int
\frac{d^3k}{(2\pi)^3}2T\ln\big{(}1+e^{\frac{+\sqrt{k^2+\mathcal{M}_i^2}}{2T}}\big{)}
.\end{align} The above is our final formula. Notice that relation
(\ref{potentialmine3golftsi}) contains integrals we computed in
the previous sections, both for bosons and for fermions, see for
example relations (\ref{24}) and (\ref{70}). Also the first term
corresponds to the scalar bosons part, the second to the gauge
bosons and the third to the fermion part. The same correspondence
applies to relation (\ref{potentialmine2golftsi}). The masses that
appear in relations (\ref{potentialmine3golftsi}) and
(\ref{potentialmine2golftsi}) are model dependent and can be found
in the same way as in (\ref{31revision}), (\ref{32arevision}) and
(\ref{42arevision}).

\noindent All the above are invaluable to the theories of phase
transitions at finite temperature. See for example reference
\cite{quirowfinite} and references therein.

\noindent In conclusion the generalization of the above to any
dimensions is straightforward. In general, apart from the phase
transition application, a theory at finite temperature offers the
possibility to connect a $d$ dimensional theory with the $d+1$
dimensional theory at finite temperature. Let us discuss a little
on this. One could say that the calculations we obtained actually
correspond to a three dimensional theory in the case of initial
$d=4$ theory. However one should be really cautious since the
argument that a $d$ dimensional field theory correspond to the
same theory in $d-1$ dimensions has been proven true \cite{largen}
only for the $\phi^4$ theory (always within the limits of
perturbation theory). Also this also holds true for supersymmetric
theories. On the contrary this does not hold for $QCD$ and
Yang-Mills theories. Actually $QCD_3$ resembles more $QCD_4$ and
not $QCD_4$ at finite temperature! It would be more correct to say
that a $d$ dimensional theory at finite temperature resembles more
the same theory with one dimension compactified to a circle and in
the limit $R\rightarrow 0$, where $R$ the magnitude of the compact
dimension. We shall report on these issues somewhere else
\cite{malakompoukoma}.

\section{Calculation of Effective Potential in Spacetime Topology $S^{1}\times R^{d}$}

In this section we will compute the fermionic and bosonic
contributions to the effective potential of field theories
quantized in spacetime topologies $S^{1}\times R^{d}$
\cite{odi1,odi2,kirstennew2,elizalde2,elizalde,kirstennew1,elizaldesp23,elizaldesp24,elizaldesp25,elizaldesp26}.
The calculations are done in Euclidean time by making a Wick
rotation in the time coordinate. By this we have static-time
independent results. In space times with non trivial topology the
fields can have periodic or antiperiodic boundary conditions
without the restrictions that we had in the temperature case
\cite{elizalde,kirsten14} (that is bosons must obey only periodic
and fermions only antiperiodic boundary conditions). We shall deal
with periodic bosons and antiperiodic fermions.

\noindent The boundary conditions for bosons are,
\begin{equation}\label{101}
\varphi (x,0)=\varphi (x,L) ,\end{equation} $L$ denoting the
compact (circle) dimension, while the fermion boundary conditions,
\begin{equation}\label{102}
\psi (x,0)=-\psi (x,L) .\end{equation}

\noindent Another more general set of boundary conditions that can
be used is the so called twisted boundary conditions of the form:
\begin{equation}\label{103}
\varphi (x,0)=e^{-iw}\varphi (x,L) ,\end{equation} for bosons and,
\begin{equation}\label{104}
\psi (x,0)=-e^{i\rho }\psi (x,L) ,\end{equation} for fermions.


\subsection{Periodic Bosons and Antiperiodic Fermions}

Using,
\begin{equation}\label{105}
\varphi (x,0)=\varphi (x,L) ,\end{equation} for bosons and,
\begin{equation}\label{106}
\psi (x,0)=-\psi (x,L) ,\end{equation} for fermions, we shall
compute the bosonic contribution,
\begin{equation}\label{107}
\frac{1}{L}\int \frac{dk^{3}}{(2\pi )^{3}}\sum_{n=-\infty }^{\infty }\ln [%
\frac{4\pi^{2} n^{2}}{L^{2}}+k^{2}+m^{2}] ,\end{equation} and also
the fermionic one,
\begin{equation}\label{108}
\frac{1}{L}\int \frac{dk^{3}}{(2\pi )^{3}}\sum_{n=-\infty }^{\infty }\ln [%
\frac{(2n+1)^{2}\pi^{2}}{L^{2}}+k^{2}+m^{2}] .\end{equation}

\noindent Following the techniques developed in the previous
sections (roughly we substitute $T\rightarrow \frac{1}{L}$),
\begin{align}\label{109}
&\frac{1}{L}\int \frac{dk^{d}}{(2\pi )^{d}}\sum_{n=-\infty }^{\infty }\ln [%
\frac{4\pi^{2} n^{2}}{L^{2}}+k^{2}+m^{2}]=\\&\notag
-\frac{1}{2}\frac{\sqrt{\pi }}{(2\pi )^{d}a}(2\pi )^{\frac{d-1}{2}}m^{d+1}\Gamma (-\nu -\frac{1}{2}+1)+\frac{1}{4%
}\frac{1}{(2\pi )^{d}}(2\pi )^{\frac{d-1}{2}}m^{d+1}\Gamma (-\nu
)\\&\notag -\frac{\sqrt{\pi }}{(2\pi )^{d}a}(2\pi
)^{\frac{d-1}{2}}m^{d+1}\Gamma (-\nu
-\frac{1}{2}+1)(a^{2})^{\frac{1}{2}-\nu }\\&\notag \times
[\sum_{l=0}^{\nu -\frac{1}{2}}\frac{((2\pi )^{2})^{\nu
-\frac{1}{2}-l}(\nu -\frac{1}{2})!}{(\nu
-\frac{1}{2}-l)!l!}(a^{2})^{l}\zeta (-2\nu  +1+2l)] ,\end{align}
for the boson case, with $\alpha=mL$ and,
\begin{align}\label{110}
&\frac{1}{L}\int \frac{dk^{d}}{(2\pi )^{d}}\sum_{n=-\infty
}^{\infty }\ln
[\frac{(2n+1)^{2}\pi^{2}}{L^{2}}+k^{2}+m^{2}]=\\&\notag
-\frac{(2\pi )^{\frac{d-1}{2}}m^{d+1}}{(2\pi )^{d}}{\Large (}
\sum\limits_{q=-\infty }^{\infty \prime}
\frac{K_{\frac{d+1}{2}}(mqL)}{(\frac{mqL}{2})^{\frac{d+1}{2}}}-\frac{1}{2}
\sum\limits_{q=-\infty }^{\infty \prime}
\frac{K_{\frac{d+1}{2}}(\frac{mqL}{2})}{(\frac{mqL}{4})^{\frac{d+1}{2}}}{\Large
)}=\\&\notag -\frac{\sqrt{\pi }}{(2\pi )^{d}a_{2}}(2\pi
)^{\frac{d-1}{2}}m^{d+1}\Gamma (-\nu -\frac{1}{2}+1)+\ \
\frac{1}{2}\frac{1}{(2\pi )^{d}}(2\pi
)^{\frac{d-1}{2}}m^{d+1}\Gamma(-\nu )\\&\notag -\frac{2\sqrt{\pi
}}{(2\pi )^{d}a_{2}}(2\pi )^{\frac{d-1}{2}}m^{d+1}\Gamma (-\nu
-\frac{1}{2}+1)(a_{2}^{2})^{\frac{1}{2}-\nu }\\&\notag \times
\Big{[}\sum_{l=0}^{\sigma }\frac{((2\pi
)^{2})^{\nu -\frac{1}{2}-l}(\nu -\frac{1}{2})!}{(\nu -\frac{1}{2}-l)!l!}(a_{2}^{2})^{l}\zeta (-2\nu  +1+2l)\Big{]} \\
&\notag +\frac{1}{2}\frac{\sqrt{\pi }}{(2\pi )^{d}a_{1}}(2\pi )^{\frac{d-1}{2}%
}m^{d+1}\Gamma (-\nu -\frac{1}{2}+1)-\ \ \frac{1}{4}\frac{1}{(2\pi
)^{d}}(2\pi )^{\frac{d-1}{2}}m^{d+1}\Gamma (-\nu )\\&\notag
+\frac{1}{2}\frac{\sqrt{\pi }}{(2\pi )^{d}a_{1}}(2\pi
)^{\frac{d-1}{2}}m^{d+1}\Gamma (-\nu
-\frac{1}{2}+1)(a_{1}^{2})^{\frac{1}{2}-\nu }\\&\notag \times
\Big{[}\sum_{l=0}^{\sigma }\frac{((2\pi )^{2})^{\nu
-\frac{1}{2}-l}(\nu -\frac{1}{2})!}{(\nu
-\frac{1}{2}-l)!l!}(a_{1}^{2})^{l}\zeta (-2\nu  +1+2l)\Big{]}
,\end{align} for the fermion case, with $\alpha _{2}=mL$ and
$\alpha_{1}=\frac{mL}{2}$.

\noindent For the case $d=3$ the bosonic contribution is:
\begin{align}\label{111}
&\frac{1}{L}\int \frac{dk^{3}}{(2\pi )^{3}}\sum_{n=-\infty }^{\infty }\ln [%
\frac{4\pi^{2} n^{2}}{L^{2}}+k^{2}+m^{2}]=\\&\notag
\frac{\frac{-m^{4}}{16\,\pi ^{2}}+\frac{m^{4}\,}{16\,\,\pi ^{2}}}{%
\varepsilon }+(\frac{\,m^{2}}{12\,L^{2}}+\frac{3\,m^{4}}{64\,\pi ^{2}}-\frac{%
\gamma \,m^{4}}{32\,\pi ^{2}}-\frac{\gamma \,m^{4}\,}{16\,L\,\pi ^{2}}-\frac{%
m^{3}}{6\,L\,\pi }-\frac{\,\pi ^{2}}{45\,L^{4}\,}+\frac{m^{4}\,\ln (2)}{%
32\,\pi ^{2}}+\\&\notag \frac{m^{4}\,\ln (2)}{32\,L\,\pi ^{2}}-\frac{m^{4}\,\ln (m)}{16\,\pi ^{2}}+%
\frac{m^{4}\,\,\ln (m)}{16\,L\,\pi ^{2}}-\frac{m^{4}\,\,\ln (L^{2}\,m^{2})%
}{32\,\,\pi ^{2}}+\frac{m^{4}\,\ln (\pi )}{32\,\pi ^{2}}+\frac{%
m^{4}\,\,\ln (\pi )}{32\,L\,\pi ^{2}}
\\&\notag -\frac{m^{4}\,\,\psi (-\left( \frac{3}{2}\right) )}{32\,\,\pi ^{2}}-%
\frac{m^{4}\,\,\psi (\frac{1}{2})}{32\,\,\pi ^{2}}+\frac{%
m^{4}\,\,\psi (\frac{5}{2})}{32\,\,\pi ^{2}}+ \\
&\notag \frac{L^{2}\,m^{6}\,\,\zeta (3)}{384\,\pi ^{4}}-\frac{L^{4}\,m^{8}\,\,%
\zeta (5)}{4096\,\pi ^{6}}) .\end{align} In equation (\ref{111})
we omitted terms of higher order in $L$. This is because we are
interested in the limit $L\rightarrow 0$.

\noindent The fermionic contribution for $d=3$ is:
\begin{align}\label{112}
&\frac{1}{L}\int \frac{dk^{3}}{(2\pi )^{3}}\sum_{n=-\infty
}^{\infty }\ln [\frac{(2n+1)^{2}\pi^{2}}{L^{2}}+k^{2}+m^{2}]=
\\&\notag \frac{\frac{-m^{4}}{16\,\pi ^{2}}+\frac{m^{4}\,}{16\,\,\pi ^{2}}}{\varepsilon }+
(-\frac{m^{2}}{6\,L^{2}}+\frac{3\,m^{4}}{64\,\pi ^{2}}
-\frac{\gamma m^{4}}{32\,\pi ^{2}}-\frac{\gamma
\,m^{4}\,}{16\,L\,\pi ^{2}}+\\&\notag \frac{14\,\,\pi
^{2}}{45\,L^{4}\,}-\frac{m^{4}\,\,\ln (L^{2}\,m^{2})}{32\,\,\pi
^{2}}+\frac{m^{4}\,\ln (\pi )}{16\,\pi ^{2}}
\\&\notag -\frac{m^{4}\,\,\psi (-\left( \frac{3}{2}\right) )}
{32\,\,\pi ^{2}}-\frac{m^{4}\,\psi (\frac{1}{2})}{32\,\,\pi ^{2}}+\frac{m^{4}\,\,\psi (\frac{5}{2})}{32\,\,\pi ^{2}} \\
&\notag +\frac{7\,\,m^{6}\,L^{2}\,\,\zeta (3)}{1536\,\pi
^{4}}-\frac{31\,L^{4}\,m^{8}\,\,\zeta (5)}{65536\,\pi ^{6}})
.\end{align}

\noindent In the case $d=2$ the bosonic contribution reads:
\begin{align}\label{113}
&\frac{1}{L}\int \frac{dk^{2}}{(2\pi )^{2}}\sum_{n=-\infty
}^{\infty }\ln [\frac{4\pi^{2}
n^{2}}{L^{2}}+k^{2}+m^{2}]=\\&\notag
(\frac{m^{2}}{4\,\sqrt{2}\,L\,\pi }+\frac{m^{3}}{6\,\sqrt{2}\,\pi }+\frac{%
m^{2}\,\ln (2)}{2\,\sqrt{2}\,L\,\pi }-\frac{m^{2}\,\ln (L^{2}\,m^{2})}{4\,%
\sqrt{2}\,L\,\pi }
\\&\notag +\frac{m^{2}\,\ln (\pi )}{2\,\sqrt{2}\,L\,\pi }-\frac{%
m^{2}\,\ln (2\,\pi )}{2\,\sqrt{2}\,L\,\pi }+\frac{\zeta ^{\prime }(-2)}{%
L^{3}}) ,\end{align} and the fermionic contribution:
\begin{align}\label{114}
&\frac{1}{L}\int \frac{dk^{2}}{(2\pi )^{2}}\sum_{n=-\infty }^{\infty }\ln [(%
\frac{(2n+1)^{2}\pi^{2}}{L^{2}}+k^{2}+m^{2}]=\\&\notag \frac{m^{3}}{6\,\sqrt{2}\,\pi }-%
\frac{m^{2}\,\ln (2)}{\sqrt{2}\,L\,\pi }-\frac{\zeta ^{\prime
}(-2)}{L^{3}} .\end{align}

\subsection{Some Applications I}

\subsubsection{Topological Symmetry Breaking in Self Interacting
Field Theories}

We now discuss some applications of the periodic bosons and
anti-periodic fermions effective potential at finite volume. It is
well known that field theory at finite volume plays an important
role to topological symmetry breaking or restoration and
topological mass generation
\cite{odi1,odi2,kirstennew2,elizalde2,elizalde,kirstennew1,elizaldesp23,elizaldesp24,elizaldesp25,elizaldesp26,elizaldesp19,spalluci,ford,ferrer,alles}.
Apart from the known influence of the topology to the boundary
conditions of the sections of the fiber bundles studied, the
effective mass and on particle creation \cite{elizalde} the need
for studying field theories at finite volume is that the universe
might exhibit non trivial topology as a whole
\cite{gongcharov,odi,odi1,elizaldesp24,elizalde}.

\noindent Now we briefly present the topological mass generation.
When spacetime has non-trivial topology then a massless field with
periodic boundary conditions, can acquire mass through loop
corrections, in a dynamical way. Indeed, the one loop potential
reads,
\begin{equation}\label{1loopmassgenerw}
V^1(\phi)=\frac{1}{vol(M)}\sum_{n}\ln(a_n/\mu^2) ,\end{equation}
with $vol(M)$ is the volume of the spacetime under study and $a_n$
are the eigenvalues of the Laplace operator on this spacetime. A
regularized form of the above involves the zeta function
\cite{kirstennew1},
\begin{equation}\label{zetapipe}
\zeta (s)=\sum_{n}a_n^{-s}.\end{equation} The potential at loop is
written as,
\begin{equation}\label{laodif}
V^1(\phi)=\frac{1}{vol(M)}[\zeta'(0)+\zeta (0)\ln\mu^2]
,\end{equation} with $\mu$ a dimensional regularization parameter
that can be removed in the renormalization process. The
topological mass is equal to,
\begin{equation}\label{topologicalarxidomass}
m^2=\frac{\mathrm{d}^2V(\phi)}{\mathrm{d}\phi^2} ,\end{equation}
at $\phi =0$. In the above relation, $V(\phi)$ is equal to,
\begin{equation}\label{tatanospasmini}
V(\phi)=\frac{\lambda}{4!}\phi^4-\frac{1}{vol(M)}[\zeta'(0)+\zeta
(0)\ln \mu^2] ,\end{equation} Now for the spacetime $S^1\times
R^3$ the eigenvalues $a_n$ are,
\begin{equation}\label{spygame1}
a_n=\frac{\lambda}{2}\phi^2+\big{(}\frac{2\pi
n}{L}+k_1^2+k_2^2+k_3^2\big{)}.\end{equation} Also the zeta
function $\zeta (s)$ reads,
\begin{equation}\label{spygame2}
\zeta (s)=\frac{L_1}{2\pi}\int
d^3k_i\sum_{n=-\infty}^{\infty}\Big{[}\frac{\lambda}{2}\phi^2+\big{(}\frac{4\pi
^2 n^2}{L^2}+k_1^2+k_2^2+k_3^2\Big{]} ,\end{equation} The
calculation of the above can be done with the techniques we
presented in the previous sections. Now at $\phi =0$ the potential
is,
\begin{equation}\label{spygame3}
V(\phi =0)=-\frac{\pi^2}{90L_1^4} ,\end{equation} The above is
just the Casimir energy for a real scalar field that satisfies
periodic boundary conditions instead of Dirichlet. The
topologically generated mass in this case is,
\begin{equation}\label{spygame4}
m^2=\frac{\lambda}{24L^2_1}.\end{equation} These techniques can be
useful to determine the vacuum stability of the theory under
consideration \cite{odi,odi1,ferrer,elizalde,ford}. In the case of
the periodic scalar field, the mass is positive, thus the $\phi
=0$ vacuum is stable. Let us now study the same setup in
$S^1\times R^3$ but with the scalar field satisfying anti-periodic
boundary conditions along the compact dimension. This case
resembles the calculations of a fermion field at finite volume we
presented previously. The only vacuum expectation value that is
allowed is $\phi =0$ \cite{isham}. The zeta function now reads,
\begin{equation}\label{spygame2}
\zeta (s)=\frac{L_1}{2\pi}\int
d^3k_i\sum_{n=-\infty}^{\infty}\Big{[}\frac{\lambda}{2}\phi^2+\big{(}\frac{\pi^2(2
n+1)^2}{L^2}+k_1^2+k_2^2+k_3^2\Big{]},\end{equation} and in this
case, at $\phi =0$ the potential is,
\begin{equation}\label{spygame3}
V(\phi =0)=\frac{7\pi^2}{720L_1^4} ,\end{equation} The above is
just the Casimir energy for a real scalar field that satisfies
periodic boundary conditions instead of Dirichlet. The
topologically generated mass now reads,
\begin{equation}\label{spygame4}
m^2=-\frac{\lambda}{48L^2_1}.\end{equation} The negative sign
indicates an instability in this theory \cite{ford,elizalde,odi}.

\subsubsection{Casimir Effect the Effective Potential and Extra Dimensions}

The calculations for finite volume field theories with a toroidal
compact dimension are useful for field theories with one compact
extra dimension. We shall present some cases here. Also these are
special cases of the effective potential with a twist in the
fields boundary conditions that we describe in the next section.

\noindent Let us start with a scalar field in the Randall-Sundrum1
(RS1) model \cite{rs1}. The line element is given by,
\begin{equation}\label{rundul}
ds^2=e^{-2kr_{c}\phi}\eta_{\mu{\,}\nu}dx^{\mu}dx^{\nu}-r_{c}^2d\phi^2
,\end{equation} The theory is quantized on the orbifold $S^1/Z_2$
and thus the points $(x^{\mu},\phi)$ and $(x^{\mu},-\phi)$ are
identified. The exponential factor is the most appealing feature
of the RS1 model. Actually the hierarchy problem can be solved
within this scenario since a Tev mass scale can be produced from a
Plank mass scale \cite{rs1}. One of the most interesting problems
appearing in models with extra compact dimensions is related with
the size and stability of the compact dimension. Particularly the
problem is two fold. First one must find a way to shrink the extra
dimensions. This is a very serious feature since the visible
spatial dimensions of our world inflated in the past. Also their
size exponentially increased during inflation. So firstly, the
extra dimensions must shrink. Second the extra dimensions must be
stabilized and not to collapse to the Plank scale. One indicator
to solve the first problem is the existence of negative energy in
the bulk, that is the Casimir energy of the bulk scalar field must
be negative. In the context of string theory there are setups such
us orientifolds planes and other structures
\cite{direstr12,direstr12}. In some cases field theory corrections
can be supplemented by string structures but we shall not discuss
this here.

\noindent Consider a free scalar in the bulk, with Lagrangian
density,
\begin{equation}\label{scalarbulk}
L=G_{A{\,}B}\partial_{A}\Phi\partial_{B}\Phi-m^2\Phi^2
,\end{equation} The harmonic expansion of the scalar field is,
\begin{equation}\label{harmonic}
\Phi(x^{\mu},\phi)=\sum_{n}\psi_{n}(x^{\mu})\frac{y_n(\phi)}{\sqrt{R}}
,\end{equation} Solving the equations of motion for the RS metric
one obtains obtain,
\begin{equation}\label{bessel}
y_n(\phi)\sim
e^{2kR\phi}\Big{[}J_{\nu}(\frac{M_ne^{kR\phi}}{k})+Y_{\nu}(\frac{M_ne^{kR\phi}}{k})\Big{]}
,\end{equation} and in order the field satisfies the orbifold
boundary conditions, $M_n$ must satisfy,
\begin{equation}\label{moni}
M_ne^{kR\phi}{k}\sim \pi(N+\frac{1}{4}) ,\end{equation} It is
clear that the Casimir energy is significant due to the extra
dimensions quantum fluctuations. For the bulk scalar field we
obtain,
\begin{equation}\label{potentialrs}
V^+=\frac{1}{2}\sum_{\nu
=-\infty}^{\infty}\frac{d^4k}{(2\pi)^4}\ln(k^2+\Big{(}\frac{n\pi}{r_c}\Big{)}^2+M_n^2)
,\end{equation} with $r_c$ the compact dimension radius. Notice
that relation (\ref{potentialrs}) is identical with relation
(\ref{107}) for the case of five dimensions. The calculation and
generalization is straightforward, and we can find the result in
closed form, in terms of the polylogarithm functions. This
calculation is similar to the finite temperature one for $d$ even,
see relations (\ref{95}) and (\ref{65}). For a more general
calculation see the next section. In the case of a massless scalar
relation (\ref{potentialrs}) is modified to,
\begin{equation}\label{potentialrs1}
V^+_{1}=\frac{1}{2}\sum_{\nu
=-\infty}^{\infty}\frac{d^4k}{(2\pi)^4}\ln(k^2+\Big{(}\frac{n\pi}{r_c}\Big{)}^2)
,\end{equation} which is calculated to be,
\begin{equation}\label{calunrs}
V^+_{1}=-\frac{3\zeta (5)}{64\pi^4r_c^4},\end{equation} which is
clearly negative, and thus this results to a shrinking of the
compact dimension. Also the Casimir force in terms of the compact
dimensions is repulsive which leads to a stabilization of the
extra dimension. The calculations for fermions are
straightforward. Also the existence of a minimum in the effective
potential is an indicator of stabilization of the extra
dimensions.

\noindent Finally let us mention that Casimir calculations have
been done for de Sitter and anti-de Sitter brane worlds, see
\cite{odintsovelizalde,odi8,odi8a,odi6}.

\noindent Additionally same results hold for other 5-dimensional
setups, such us large extra dimensions and universal extra
dimensions. We shall briefly present some applications in relation
to them after the next section.

\subsection{The Case of Twisted Boundary Conditions}

We shall study only the twisted boson case since the other case is
similar \cite{elizalde,kirstennew1,elizalde2}.

\noindent The twisted boundary conditions for bosons are:
\begin{equation}\label{115}
\varphi (x,0)=e^{-iw}\varphi (x,L) ,\end{equation} while for
fermions:
\begin{equation}\label{116}
\psi (x,0)=-e^{i\rho }\psi (x,L) ,\end{equation} or equivalently,
\begin{equation}\label{117}
\psi (x,0)=e^{i(\rho +\pi )}\psi (x,L) .\end{equation}

\noindent We Fourier expand $\varphi $:
\begin{equation}\label{118}
\sum\limits_{n}\int dp^{3}e^{ipx}=e^{iw}\sum\limits_{n}\int
dp^{3}e^{ipx+iw_{n}L}
,\end{equation}%
from which we obtain,
\begin{equation}\label{119}
w_{n}L=2\pi n+w\rightarrow w_{n}=(2\pi n+w)\frac{1}{L}
,\end{equation} with, $G=\frac{1}{w_{n}^{2}+k^{2}+m^{2}}$.

\noindent Doing the same as in the previous with the difference:
\begin{equation}\label{120}
w_{n}=(2\pi n+w)\frac{1}{L}=(n+\omega )\frac{2\pi }{L}
,\end{equation} with, $\omega $=$\frac{w}{2\pi }$, we will compute
\cite{elizalde,kirstennew1,elizalde2},
\begin{equation}\label{121}
\frac{1}{L}\int \frac{dk^{3}}{(2\pi )^{3}}\sum \ln [((n+\omega
)\frac{2\pi }{L})^{2}+k^{2}+m^{2}] .\end{equation}

\noindent Consider the sum:
\begin{equation}\label{122}
\sum\limits_{n=-\infty }^{\infty }\frac{1}{(n+\omega )^{2}(\frac{2\pi }{L}%
)^{2}+a^{2}}=\frac{1}{(\frac{2\pi }{L})^{2}}\sum\limits_{n=-\infty
}^{\infty }\frac{1}{(n+\omega )^{2}+\frac{a^{2}}{(2\pi
\frac{1}{L})^{2}}} ,\end{equation} with,
\begin{equation}\label{123}
a^{2}=k^{2}+m^{2} .\end{equation} Integrating,
\begin{equation}\label{124}
\sum\limits_{n=-\infty }^{\infty }\frac{1}{(n+\omega )^{2}(\frac{2\pi }{L}%
)^{2}+a^{2}} ,\end{equation} over $a^{2}$, we get,
\begin{equation}\label{125}
\ \int \ \sum\limits_{n=-\infty }^{\infty }\frac{da^{2}}{(n+\omega )^{2}(%
\frac{2\pi }{L})^{2}+a^{2}}=\sum_{n=-\infty }^{\infty }\ln [(n+\omega )^{2}(%
\frac{2\pi }{L})^{2}+a^{2}] .\end{equation} Also,
\begin{equation}\label{126}
\sum\limits_{n=-\infty }^{\infty }\frac{da^{2}}{(n+\omega
)^{2}(\frac{2\pi
}{L})^{2}+a^{2}}=\frac{L}{4a}\Big{(}\coth (\frac{aL}{2}-i\pi \omega )+\coth (\frac{aL}{%
2}+i\pi \omega )\Big{)} ,\end{equation} and consequently,
\begin{align}\label{126}
&\int \sum\limits_{n=-\infty }^{\infty }\frac{da^{2}}{(n+\omega )^{2}(%
\frac{2\pi }{L})^{2}+a^{2}}=\\&\nobreak \int \frac{L}{4a}(\coth
(\frac{aL}{2}-i\pi \omega )+\coth (\frac{aL}{2}+i\pi \omega
)da^{2}=\\&\notag \ln (\sinh [\frac{aL}{2}-i\pi \omega ])+\ln
(\sinh [\frac{aL}{2}+i\pi \omega )) .\end{align} Using
\cite{qradstein7},
\begin{equation}\label{127}
\ln (\sinh x)=\ln (\frac{1}{2}(e^{x}-e^{-x}))=x+\ln
(1-e^{-2x})-\ln [2] ,\end{equation} and summing,
\begin{equation}\label{128}
\ln (\sinh  [\frac{aL}{2}-i\pi \omega ])=\frac{aL}{2}-i\pi \omega +\ln [1-e^{-2(%
\frac{aL}{2}-i\pi \omega )}]-\ln [2] ,\end{equation} and,
\begin{equation}\label{129}
\ln (\sinh  [\frac{aL}{2}+i\pi \omega ])=\frac{aL}{2}+i\pi \omega +\ln [1-e^{-2(%
\frac{aL}{2}+i\pi \omega )}]-\ln [2] .\end{equation} we get,
\begin{align}\label{130}
&\int \sum\limits_{n=-\infty }^{\infty }\frac{da^{2}}{(n+\omega )^{2}(%
\frac{2\pi }{L})^{2}+a^{2}}=\\&\notag \ln (\sinh
[\frac{aL}{2}-i\pi \omega])+\ln (\sinh [\frac{aL}{2}+i\pi \omega
))=\\&\notag aL+\ln [1-e^{-2(\frac{aL}{2}-i\pi \omega )}]+\ln
[1-e^{-2(\frac{aL}{2}+i\pi \omega )}]-2\ln [2] .\end{align} After
some calculations
\cite{elizalde,kirstennew1,elizalde2,elizaldesp1}:
\begin{align}\label{skata}
&\sum_{n=-\infty }^{\infty }\ln [(n+\omega )^{2}(\frac{2\pi }{L}%
)^{2}+a^{2}]=\\&\notag \alpha L+\ln [1-e^{-2(\frac{aL}{2}-i\pi \omega )}]+\ln [1-e^{-2(%
\frac{aL}{2}+i\pi \omega )}]-2\ln [2]\ .\end{align} Using the
identity \cite{qradstein7},
\begin{equation}\label{131}
\sum \ln [\frac{(n+\omega )^{2}4\pi^{2}T^{2}+a^{2})}{(n+\omega
)^{2}4\pi^{2}T^{2}+b^{2})}]=2(a-b)\ \ ,\end{equation} the relation
(\ref{skata}) becomes,
\begin{align}\label{132}
&\sum \ln [(n+\omega )^{2}(\frac{2\pi }{L})^{2}+a^{2}]=\\&\notag \frac{L}{2\pi }%
\int\limits_{-\infty }^{\infty }dx\ln [x^{2}+a^{2}]+\ln [1-e^{-2(\frac{aL}{2}%
-i\pi \omega )}]+\ln [1-e^{-2(\frac{aL}{2}+i\pi \omega )}]
.\end{align} Thus,
\begin{align}\label{mouni}
&\frac{1}{L}\int \frac{dk^{3}}{(2\pi )^{3}}\sum \ln [((n+\omega )\frac{2\pi }{L}%
)^{2}+k^{2}+m^{2}]=\\&\notag \int \frac{dk^{3}}{(2\pi
)^{3}}\int\limits_{-\infty }^{\infty }\frac{dx}{2\pi }\ln
[x^{2}+a^{2}] \\&\notag +\frac{1}{L}\int \frac{dk^{3}}{(2\pi
)^{3}}\ln [1-e^{-2(\frac{aL}{2}-i\pi
\omega )}]\\&\notag +\frac{1}{L}\int \frac{dk^{3}}{(2\pi )^{3}}\ln [1-e^{-2(\frac{aL}{2%
}+i\pi \omega )}] \nonumber ,\end{align} with,
\begin{equation}\label{133}
a^{2}=k^{2}+m^{2} ,\end{equation} The first integral is the one
loop correction to the effective potential for $L=0$. In $d+1$
dimensions relation (\ref{mouni}) reads
\cite{elizalde,elizalde2,kirstennew1,elizaldesp25,Milton3,referee1,referee2}:
\begin{align}\label{134}
&\frac{1}{L}\int \frac{dk^{d}}{(2\pi )^{d}}\sum \ln [((n+\omega
)\frac{2\pi
}{L})^{2}+k^{2}+m^{2}]=\\&\notag \int \frac{dk^{d+1}}{(2\pi )^{d+1}}%
\ln [k^{2}+a^{2}]\\&\notag +\frac{1}{L}\int \frac{dk^{d}}{(2\pi
)^{d}}\ln [1-e^{-2(\frac{aL}{2}-i\pi
\omega )}]+\frac{1}{L}\int \frac{dk^{d}}{(2\pi )^{d}}\ln [1-e^{-2(\frac{aL}{2%
}+i\pi \omega )}] .\end{align}

\noindent In the following we consider only the $L$ dependent
part,
\begin{equation}\label{135}
V_{twisted}=\frac{1}{L}\int \frac{dk^{d}}{(2\pi )^{d}}\ln [1-e^{-2(\frac{aL}{2}%
-i\pi \omega )}]+\frac{1}{L}\int \frac{dk^{d}}{(2\pi )^{d}}\ln [1-e^{-2(%
\frac{aL}{2}+i\pi \omega )}] .\end{equation}

\noindent Let,
\begin{equation}\label{136}
V_{1}=\frac{1}{L}\int \frac{dk^{d}}{(2\pi )^{d}}\ln [1-e^{-2(\frac{aL}{2}%
-i\pi \omega )}] ,\end{equation} and
\begin{equation}\label{137}
V_{2}=\frac{1}{L}\int \frac{dk^{d}}{(2\pi )^{d}}\ln [1-e^{-2(\frac{aL}{2}%
+i\pi \omega )}] .\end{equation} so relation (\ref{135}) reads,

\begin{equation}\label{bourdelo}
V_{twisted}=V_{1}+V_{2} ,\end{equation}

\noindent The calculation of $V_{1}$ and of $V_{2}$ is equivalent.
Their analytic properties are the same. So we calculate only
$V_{2}$. We have,
\begin{equation}\label{138}
V_{2}=\frac{1}{L}\int \frac{dk^{d}}{(2\pi )^{d}}\ln [1-e^{-2(\frac{aL}{2}%
+i\pi \omega )}]=\frac{1}{L}\int \frac{dk^{d}}{(2\pi )^{d}}%
\ln [1-e^{-aL-2i\pi \omega )}] .\end{equation} Using,
\begin{equation}\label{139}
\ln[1-e^{-aL-i2\pi \omega }]=-\sum\limits_{q=1}^{\infty
}\frac{e^{-aLq-2\pi i\omega q}}{q} .\end{equation} Now $V_{2}$
becomes,
\begin{align}\label{140}
&V_{2}=\frac{1}{L}\int \frac{dk^{d}}{(2\pi )^{d}}\ln
[1-e^{-aL-2i\pi \omega
)}]\\&\notag =-\frac{1}{L}\int \frac{dk^{d}}{(2\pi )^{d}}\sum\limits_{q=1}^{\infty }%
\frac{e^{-aLq-2\pi i\omega q}}{q}\\&\notag =-\sum\limits_{q=1}^{\infty }\frac{1}{L}%
\int \frac{dk^{d}}{(2\pi )^{d}}\frac{e^{-aLq-2\pi i\omega
q}}{q}\\&\notag =-\sum\limits_{q=1}^{\infty }\frac{1}{L}\int
\frac{dk^{d}}{(2\pi )^{d}}
\frac{e^{-\sqrt{k^{2}+m^{2}}qL-2\pi i\omega q}}{q}\\&\notag =-\sum\limits_{q=1}^{%
\infty }\frac{1}{L}\int_{-\infty }^{\infty }\frac{dk}{(2\pi )^{d}}k^{d-1}%
\frac{(2\pi )^{\frac{d}{2}}}{\Gamma (\frac{d}{2})}\frac{e^{-\sqrt{k^{2}+m^{2}%
}qL}}{q}e^{-2\pi i\omega q} \\&\notag =-\sum\limits_{q=1}^{\infty
}\frac{1}{L}\frac{(2\pi )^{\frac{d}{2}}}{\Gamma
(\frac{d}{2})q(2\pi )^{d}}\int_{-\infty }^{\infty }dkk^{d-1}e^{-\sqrt{%
k^{2}+m^{2}}qL}e^{-2\pi i\omega q} ,\end{align} we used
($a=\sqrt{k^{2}+m^{2}}$). The integral,
\begin{equation}\label{141}
\int_{-\infty }^{\infty }dkk^{d-1}e^{-\sqrt{k^{2}+m^{2}}qL}
,\end{equation} equals to \cite{qradstein7},
\begin{equation}\label{142}
\int_{-\infty }^{\infty }dkk^{d-1}e^{-\sqrt{k^{2}+m^{2}}qL}=2^{\frac{d}{2}%
-1}(\sqrt{\pi })^{-1}(qL)^{\frac{1}{2}-\frac{d}{2}}m^{\frac{d+1}{2}}\Gamma (%
\frac{d}{2})K_{\frac{d+1}{2}}(mqL) .\end{equation} thus $V_2$ is
written:
\begin{eqnarray}\label{143}
V_{2} &=&-\sum\limits_{q=1}^{\infty }\frac{2^{\frac{d}{2}-1}}{(2\pi )^{d}}%
(2\pi )^{\frac{d+1}{2}}m^{d+1}\frac{K_{\frac{d+1}{2}}(mqL)}{1}(\frac{1}{mqL})^{%
\frac{d+1}{2}}e^{-2\pi i\omega q} \\
&=&-\frac{1}{2}\sum\limits_{q=1}^{\infty }\frac{1}{(2\pi )^{d}}(2\pi )^{%
\frac{d-1}{2}}m^{d+1}\frac{K_{\frac{d+1}{2}}(mqL)}{(\frac{mqL}{2})^{\frac{d+1}{2}}}%
e^{-2\pi i\omega q}  \nonumber .\end{eqnarray} Equivalently $V_1$
equals to:
\begin{equation}\label{144}
V_{1}=-\frac{1}{2}\sum\limits_{q=1}^{\infty }\frac{1}{(2\pi )^{d}}(2\pi )^{%
\frac{d-1}{2}}m^{d+1}\frac{K_{\frac{d+1}{2}}(mqL)}{(\frac{mqL}{2})^{\frac{d+1}{2}}}%
e^{+2\pi i\omega q} .\end{equation} Summing $V_1$ and $V_2$
\begin{equation}\label{145}
V_{1}+V_{2}=-\frac{1}{2}\sum\limits_{q=1}^{\infty }\frac{1}{(2\pi )^{d}}%
(2\pi )^{\frac{d-1}{2}}m^{d+1}\frac{K_{\frac{d+1}{2}}(mqL)}{(\frac{mqL}{2})^{\frac{%
d+1}{2}}}(e^{+2\pi i\omega q}+e^{-2\pi i\omega q}) .\end{equation}
and using,
\begin{equation}\label{146}
\cos x=\frac{1}{2}(e^{-ix}+e^{ix}) ,\end{equation} we get:
\begin{equation}\label{gamiseta}
V_{1}+V_{2}=-\sum\limits_{q=1}^{\infty }\frac{1}{(2\pi )^{d}}(2\pi )^{\frac{%
d-1}{2}}m^{d+1}\frac{K_{\frac{d+1}{2}}(mqL)}{(\frac{mqL}{2})^{\frac{d+1}{2}}}%
\cos (2\pi \omega q) .\end{equation} The function,
\begin{equation}\label{147}
\frac{K_{\frac{d+1}{2}}(mqL)}{(\frac{mqL}{2})^{\frac{d+1}{2}}}\cos
(2\pi \omega q) ,\end{equation} is invariant under the
transformation $q\rightarrow -q$ and relation (\ref{gamiseta}) is
written,
\begin{equation}\label{148}
V_{1}+V_{2}=-\sum\limits_{q=1}^{\infty }\frac{1}{(2\pi )^{d}}(2\pi
)^{\frac{d-1}{2}}m^{d+1}\frac{K_{\frac{d+1}{2}}(mqL)}{(\frac{mqL}{2})^{\frac{d+1}{2}}}\cos
(2\pi \omega q) ,\end{equation} and finally,
\begin{equation}\label{149}
V_{1}+V_{2}=-\frac{1}{2}\sum\limits_{q=-\infty }^{\infty \prime
\prime }\frac{1}{(2\pi
)^{d}}(2\pi )^{\frac{d-1}{2}}m^{d+1}\frac{K_{\frac{d+1}{2}}(mqL)}{(\frac{mqL}{2})^{%
\frac{d+1}{2}}}\cos (2\pi \omega q) .\end{equation} Again the
symbol $ ^\prime$ means omission of the zero modes.

\noindent By breaking the cosine function to exponentials, we
introduce $F_1$ and $F_{2}$ with $V_{twist}=F_{1}+F_{2}$, where,
\begin{equation}\label{150}
F_{1}=-\frac{1}{4}\sum\limits_{q=-\infty }^{\infty \prime }\frac{1}{%
(2\pi )^{d}}(2\pi )^{\frac{d-1}{2}}m^{d+1}\frac{K_{\frac{d+1}{2}}(mqL)}{(\frac{mqL%
}{2})^{\frac{d+1}{2}}}e^{-2\pi i\omega q} ,\end{equation} and,
\begin{equation}\label{151}
F_{2}=-\frac{1}{4}\sum\limits_{q=-\infty }^{\infty \prime }\frac{1}{%
(2\pi )^{d}}(2\pi )^{\frac{d-1}{2}}m^{d+1}\frac{K_{\frac{d+1}{2}}(mqL)}{(\frac{mqL%
}{2})^{\frac{d+1}{2}}}e^{2\pi i\omega q} .\end{equation}

\noindent We compute $F_{1}$ only, since the computation of the
other is similar. We have:
\begin{equation}\label{152}
\frac{K_{\nu }(z)}{(\frac{z}{2})^{\nu }}=\frac{1}{2}\int\limits_{0}^{\infty }%
\frac{e^{-t-\frac{z^{2}}{4t}}}{t^{\nu +1}}dt ,\end{equation} and
$F_1$ becomes:
\begin{equation}\label{153}
F_{1}=-\frac{1}{8}\frac{1}{(2\pi )^{d}}(2\pi )^{\frac{d-1}{2}%
}m^{d+1}\int_{0}^{\infty }e^{-t}\frac{\sum\limits_{q=-\infty
}^{\infty
\prime }e^{-\frac{(mqL)^{2}}{4t}}e^{-2\pi i\omega q}}{t^{\frac{d+1}{2}%
+1}} .\end{equation}
 Using the Poisson identity \cite{elizalde,elizalde2,kirstennew1},
\begin{equation}\label{154}
\sum_{n=-\infty }^{\infty }f(n)=\sum_{k=-\infty }^{\infty
}\int_{-\infty }^{\infty }f(x_{1})e^{-2\pi ikx_{1}}dx_{1}
,\end{equation} with,
\begin{equation}\label{155}
f(x)=e^{-\frac{(mxL)^{2}}{4t}}e^{-2\pi i\omega x} ,\end{equation}
and $\lambda =\frac{(mL)^{2}}{4t}$, $\beta =2 $, $\pi \omega $, we
get \cite{kirstennew3}:
\begin{align}\label{156}
&\sum_{q=-\infty }^{\infty }e^{-\lambda q^{2}}e^{-i\beta q}
=\\&\notag \sum_{k=-\infty }^{\infty }\int_{-\infty }^{\infty
}e^{-\lambda x^{2}}e^{-i\beta x}e^{-2\pi ikx}dx=\\&\notag
\sqrt{2\pi }\sum_{k=-\infty }^{\infty }\frac{1}{\sqrt{2\pi
}}\int_{-\infty }^{\infty }e^{-\lambda x^{2}}e^{-i\beta x}e^{-2\pi
ikx}dx =\\&\notag \sqrt{2\pi } \sum_{k=-\infty }^{\infty
}\frac{1}{\sqrt{2\pi }}\int_{-\infty }^{\infty }e^{-\lambda
x^{2}}e^{ix(-\beta -2\pi k)}dx .\end{align} The Fourier
transformation of the function $e^{-\lambda x^{2}}$ is:
\begin{equation}\label{157}
\frac{1}{\sqrt{2\pi }}\int_{-\infty }^{\infty }e^{-\lambda
x^{2}}e^{ix(-\beta -2\pi ikx)}dx=\frac{e^{-\frac{(\beta +2\pi k)^{2}}{%
4\lambda }}}{\sqrt{2}\sqrt{\lambda }} ,\end{equation} and finally,
\begin{align}\label{158}
&\sum_{q=-\infty }^{\infty }e^{-\lambda q^{2}}e^{-i\beta
q}=\sum_{k=-\infty
}^{\infty }\sqrt{2\pi }\frac{e^{-\frac{(\beta +2\pi k)^{2}}{4\lambda }}}{%
\sqrt{2}\sqrt{\lambda }}=\\&\notag\sum_{k=-\infty }^{\infty }\sqrt{\pi }\frac{e^{-%
\frac{(\beta +2\pi k)^{2}}{4\lambda }}}{\sqrt{\lambda }}=\sqrt{\frac{\pi }{%
\lambda }}\sum_{k=-\infty }^{\infty }e^{-\frac{(\beta +2\pi
k)^{2}}{4\lambda }} .\end{align} Neglecting the zero modes we get:
\begin{equation}\label{159}
\sum_{q=-\infty }^{\infty }e^{-\lambda q^{2}}e^{-i\beta q}=\sqrt{\frac{\pi }{%
\lambda }}\sum_{k=-\infty }^{\infty }e^{-\frac{(\beta +2\pi
k)^{2}}{4\lambda }} ,\end{equation} from which,
\begin{equation}\label{160}
1+\sum_{q=-\infty }^{\infty \prime}e^{-\lambda q^{2}}e^{-i\beta q}=\sqrt{%
\frac{\pi }{\lambda }}(\ \ e^{-\frac{\beta ^{2}}{4\lambda }}\ +\
\sum_{k=-\infty}^{\infty \prime}e^{-\frac{(\beta +2\pi
k)^{2}}{4\lambda }}) ,\end{equation} or equivalently,
\begin{equation}\label{161}
\ \sum_{q=-\infty }^{\infty \prime}e^{-\lambda q^{2}}e^{-i\beta q}=\sqrt{\frac{%
\pi }{\lambda }}(\ e^{-\frac{\beta ^{2}}{4\lambda }}\ +\
\sum_{k=-\infty}^{\infty \prime}e^{-\frac{(\beta +2\pi
k)^{2}}{4\lambda }})\ -\ 1 .\end{equation}

\noindent Replacing in $F_{1}$ we obtain,
\begin{equation}\label{162}
F_{1}=-\frac{1}{8}\frac{1}{(2\pi )^{d}}(2\pi )^{\frac{d-1}{2}%
}m^{d+1}\int_{0}^{\infty }dte^{-t}(\frac{\sqrt{\frac{\pi }{\lambda }}(\ e^{-%
\frac{\beta ^{2}}{4\lambda }}\ +\ \sum_{k=-\infty}^{\infty \prime}e^{-\frac{%
(\beta +2\pi k)^{2}}{4\lambda }})\ -\ 1}{t^{\frac{d+1}{2}+1}})
.\end{equation}

\noindent Setting, \begin{equation}\label{pordei} v=\frac{d+1}{2}
,\end{equation} and the above becomes,
\begin{eqnarray}\label{163}
F_{1} &=&-\frac{1}{8}\frac{1}{(2\pi )^{d}}(2\pi )^{\frac{d-1}{2}%
}m^{d+1}(\int_{0}^{\infty }dte^{-t}\frac{\sqrt{\frac{\pi }{\lambda }}e^{-%
\frac{\beta ^{2}}{4\lambda }}\ }{t^{\nu +1}}) \\
&&-\frac{1}{8}\frac{1}{(2\pi )^{d}}(2\pi )^{\frac{d-1}{2}}m^{d+1}\int_{0}^{%
\infty }dte^{-t}(\frac{\sqrt{\frac{\pi }{\lambda }}(\
\sum_{k=-\infty}^{\infty \prime}e^{-\frac{(\beta +2\pi k)^{2}}{4\lambda }})\ }{t^{\nu +1}})\nonumber \\
&&+\frac{1}{8}\frac{1}{(2\pi )^{d}}(2\pi )^{\frac{d-1}{2}}m^{d+1}\int_{0}^{%
\infty }dte^{-t}(\frac{1}{t^{\nu +1}})  \nonumber .\end{eqnarray}
Substitute $a=mL$ and the above relation is written $(\lambda
=\frac{a^{2}}{4t})$,
\begin{align}\label{164}
&F_{1}=-\frac{1}{8}\frac{1}{(2\pi )^{d}}(2\pi )^{\frac{d-1}{2}%
}m^{d+1}(\int_{0}^{\infty }dte^{-t}\frac{\sqrt{\pi t}2e^{-\frac{\beta ^{2}}{%
a^{2}}t}\ }{at^{\nu +1}})\\&\notag
-\frac{1}{8}\frac{1}{(2\pi )^{d}}(2\pi )^{\frac{d-1}{2}}m^{d+1}\int_{0}^{%
\infty }dte^{-t}(\frac{\sqrt{\pi t}2(\ \sum_{k=-\infty}^{\infty \prime}e^{-\frac{%
(\beta +2\pi k)^{2}}{a^{2}}t})\ }{at^{\nu +1}})\\&\notag +\frac{1}{8}\frac{1}{(2\pi )^{d}}%
(2\pi )^{\frac{d-1}{2}}m^{d+1}\int_{0}^{\infty
}dte^{-t}(\frac{1}{t^{\nu +1}}) .\end{align} After some
calculations we get:
\begin{align}\label{165}
&F_{1}=-\frac{1}{4}\frac{\sqrt{\pi }}{(2\pi )^{d}a}(2\pi )^{\frac{d-1}{2}%
}m^{d+1}(\int_{0}^{\infty }dte^{-(\frac{\beta ^{2}}{a^{2}}+1)t}\ t^{-\nu -\frac{%
1}{2}})\\&\notag -\frac{1}{4}\frac{\sqrt{\pi }}{(2\pi )^{d}a}(2\pi )^{\frac{d-1}{2}%
}m^{d+1}\int_{0}^{\infty }dte^{-t}(\frac{\sqrt{\pi t}2(\
\sum_{k=-\infty}^{\infty \prime}e^{-\frac{(\beta +2\pi
k)^{2}}{a^{2}}t})\ }{at^{\nu +\frac{1}{2}}})
\\&\notag +\frac{1}{8}\frac{1}{(2\pi )^{d}}(2\pi )^{\frac{d-1}{2}}m^{d+1}\int_{0}^{%
\infty }dte^{-t}(\frac{1}{t^{\nu +1}}) .\end{align} Finally using
the following,
\begin{equation}\label{wedge}
\frac{1}{(x^{2}+a^{2})^{\mu +1}}=\frac{1}{\Gamma (\mu
+1)}\int_{0}^{\infty }dte^{-(x^{2}+a^{2})t}t^{\mu }
,\end{equation} we have:
\begin{eqnarray}\label{166}
F_{1} &=&-\frac{1}{4}\frac{\sqrt{\pi }}{(2\pi )^{d}a}(2\pi )^{\frac{d-1}{2}%
}m^{d+1}\Gamma (-\nu -\frac{1}{2}+1)(\frac{\beta ^{2}}{a^{2}}+1)^{\nu +\frac{1}{2}%
-1}\nonumber \\
&&-\frac{1}{4}\frac{\sqrt{\pi }}{(2\pi )^{d}a}(2\pi )^{\frac{d-1}{2}%
}m^{d+1}\Gamma (-\nu -\frac{1}{2}+1)[\sum_{k=-\infty}^{\infty \prime}(1+(\frac{%
\beta +2\pi k}{a})^{2})^{\nu +\frac{1}{2}-1}]\nonumber \\
&& +\frac{1}{8}\frac{1}{(2\pi )^{d}}(2\pi
)^{\frac{d-1}{2}}m^{d+1}\Gamma (-\nu ) .\end{eqnarray}

\noindent Adding F$_{2}$ (with $-\beta +2\pi k$) we have,
\begin{align}\label{167}
&V_{twist}=-\frac{1}{2}\frac{\sqrt{\pi }}{(2\pi )^{d}a}(2\pi )^{\frac{d-1}{2}%
}m^{d+1}\Gamma (-\nu -\frac{1}{2}+1)(\frac{\beta ^{2}}{a^{2}}+1)^{\nu +\frac{1}{2}%
-1}\\&\notag
-\frac{1}{4}\frac{\sqrt{\pi }}{(2\pi )^{d}a}(2\pi )^{\frac{d-1}{2}%
}m^{d+1}\Gamma (-\nu -\frac{1}{2}+1)\\&\notag \times [\sum_{k=-\infty}^{\infty \prime}(1+(\frac{%
\beta +2\pi k}{a})^{2})^{\nu +\frac{1}{2}-1}+(1+(\frac{-\beta +2\pi k}{a}%
)^{2})^{\nu +\frac{1}{2}-1}]\\&\notag +\frac{1}{4}\frac{1}{(2\pi
)^{d}}(2\pi )^{\frac{d-1}{2}}m^{d+1}\Gamma (-\nu ) .\end{align}
The sum,
\begin{equation}\label{168}
\sum_{k=-\infty}^{\infty \prime}(1+(\frac{\beta +2\pi k}{a})^{2})^{\nu +\frac{1}{2}%
-1}+(1+(\frac{-\beta +2\pi k}{a})^{2})^{\nu +\frac{1}{2}-1}
,\end{equation}

\noindent is invariant under $k\rightarrow-k$, thus:
\begin{equation}\label{169}
2\sum_{k=1}^{\infty }(1+(\frac{\beta +2\pi k}{a})^{2})^{\nu
+\frac{1}{2}-1}+(1+(\frac{-\beta +2\pi k}{a})^{2})^{\nu
+\frac{1}{2}-1} ,\end{equation} So we obtain:
\begin{align}\label{170}
&V_{twist}=-\frac{1}{2}\frac{\sqrt{\pi }}{(2\pi )^{d}a}(2\pi
)^{\frac{d-1}{2}}m^{d+1}\Gamma (-\nu -\frac{1}{2}+1)(\frac{\beta
^{2}}{a^{2}}+1)^{\nu +\frac{1}{2}-1}\\&\notag
-\frac{1}{2}\frac{\sqrt{\pi }}{(2\pi )^{d}a}(2\pi
)^{\frac{d-1}{2}}m^{d+1}\Gamma (-\nu
-\frac{1}{2}+1)(a^{2})^{\frac{1}{2}-\nu }\\&\notag \times
[\sum_{k=1}^{\infty }(a^{2}+(\beta +2\pi k)^{2})^{\nu
+\frac{1}{2}-1}+(a^{2}+(-\beta +2\pi k)^{2})^{\nu
+\frac{1}{2}-1}]\\&\notag +\frac{1}{4}\frac{1}{(2\pi )^{d}}(2\pi
)^{\frac{d-1}{2}}m^{d+1}\Gamma (-\nu ) .\end{align} Depending on
whether $d$ is even or odd we can Taylor expand or use the
binomial expansion for the sum \cite{qradstein7}:
\begin{equation}\label{171}
(a^{2}+b^{2})^{\nu -\frac{1}{2}}=\sum_{l=0}^{\sigma }\frac{(\nu -\frac{1}{2})!}{(\nu -%
\frac{1}{2}-l)!l!}(a^{2})^{l}(b^{2})^{\nu -\frac{1}{2}-l}
.\end{equation} If $d$ is even then $\sigma =\nu -\frac{1}{2}$. If
$d$ is odd, then $\sigma $ is a positive integer.

\noindent For $d$ odd, we Taylor expand:
\begin{align}\label{172}
&V_{twist}=-\frac{1}{2}\frac{\sqrt{\pi }}{(2\pi )^{d}a}(2\pi )^{\frac{d-1}{2}%
}m^{d+1}\Gamma (-\nu -\frac{1}{2}+1)(\frac{\beta ^{2}}{a^{2}}+1)^{\nu +\frac{1}{2}%
-1}+\\&\notag \frac{1}{4}\frac{1}{(2\pi )^{d}}(2\pi
)^{\frac{d-1}{2}}m^{d+1}\Gamma (-\nu )\\&\notag
-\frac{1}{2}\frac{\sqrt{\pi }}{(2\pi )^{d}a}(2\pi )^{\frac{d-1}{2}%
}m^{d+1}\Gamma (-\nu -\frac{1}{2}+1)(a^{2})^{\frac{1}{2}-\nu
}\\&\notag \times [\sum_{k=1}^{\infty
}\sum_{l=0}^{\sigma }\frac{(\nu -\frac{1}{2})!}{(\nu -\frac{1}{2}-l)!l!}%
(a^{2})^{l}((\beta +2\pi k)^{2})^{\nu -\frac{1}{2}-l}\\&\notag
+\sum_{k=1}^{\infty }\sum_{l=0}^{\sigma }\frac{(\nu -\frac{1}{2})!}{(\nu -\frac{1}{%
2}-l)!l!}(a^{2})^{l}((-\beta +2\pi k)^{2})^{\nu -\frac{1}{2}-l}]\
,\end{align} and after calculations,
\begin{align}\label{173}
&V_{twist}=-\frac{1}{2}\frac{\sqrt{\pi }}{(2\pi )^{d}a}(2\pi )^{\frac{d-1}{2}%
}m^{d+1}\Gamma (-\nu -\frac{1}{2}+1)(\frac{\beta ^{2}}{a^{2}}+1)^{\nu +\frac{1}{2}%
-1}+\\&\notag \frac{1}{4}\frac{1}{(2\pi )^{d}}(2\pi
)^{\frac{d-1}{2}}m^{d+1}\Gamma (-\nu )\\&\notag
-\frac{1}{2}\frac{\sqrt{\pi }}{(2\pi )^{d}a}(2\pi )^{\frac{d-1}{2}%
}m^{d+1}\Gamma (-\nu -\frac{1}{2}+1)(a^{2})^{\frac{1}{2}-\nu }((2\pi )^{2})^{\nu -%
\frac{1}{2}-l}\\&\notag \times [\sum_{k=1}^{\infty }\sum_{l=0}^{\sigma }\frac{(\nu -\frac{1}{2})!%
}{(\nu -\frac{1}{2}-l)!l!}(a^{2})^{l}((\frac{\beta }{2\pi }+k)^{2})^{\nu -\frac{1}{%
2}-l}\\&\notag
+\sum_{k=1}^{\infty }\sum_{l=0}^{\sigma }\frac{(\nu -\frac{1}{2})!}{(\nu -\frac{1}{%
2}-l)!l!}(a^{2})^{l}((-\frac{\beta }{2\pi }+k)^{2})^{\nu
-\frac{1}{2}-l}] .\end{align} We use zeta regularization,
expressed in terms of the Hurwitz zeta
\cite{elizalde,elizalde2,kirstennew1,tich,elizaldenew1,kirsten14}:
\begin{equation}\label{174}
\zeta (s,\upsilon )=\sum_{k=0}^{\infty }\frac{1}{(k+\upsilon )^{s}}%
\rightarrow\sum_{k=1}^{\infty }\frac{1}{(k+\upsilon )^{s}}=\zeta (s,\upsilon )-\frac{1%
}{\upsilon ^{s}} .\end{equation}

\noindent which is defined for $0<\upsilon \leq 1$ and the term
$k+\upsilon =0$ is omitted. In our case $\upsilon $ is $\beta $
which contains the phase appearing in the boundary conditions. So
$\omega$ must be positive ($\beta =\frac{\omega }{2\pi }$).

\noindent Using Hurwitz zeta
\cite{elizalde,elizalde2,kirstennew1,tich,elizaldenew1,kirsten14}:
\begin{align}\label{175}
&V_{twist}=-\frac{1}{2}\frac{\sqrt{\pi }}{(2\pi )^{d}a}(2\pi )^{\frac{d-1}{2}%
}m^{d+1}\Gamma (-\nu -\frac{1}{2}+1)(\frac{\beta ^{2}}{a^{2}}+1)^{\nu +\frac{1}{2}%
-1}\\&\notag +\frac{1}{4}\frac{1}{(2\pi )^{d}}(2\pi
)^{\frac{d-1}{2}}m^{d+1}\Gamma (-\nu )\\&\notag
-\frac{1}{2}\frac{\sqrt{\pi }}{(2\pi )^{d}a}(2\pi )^{\frac{d-1}{2}%
}m^{d+1}\Gamma (-\nu -\frac{1}{2}+1)(a^{2})^{\frac{1}{2}-\nu }((2\pi )^{2})^{\nu -%
\frac{1}{2}-l}\\&\notag \times[\sum_{k=1}^{\infty }\sum_{l=0}^{\sigma }\frac{(\nu -\frac{1}{2})!%
}{(\nu -\frac{1}{2}-l)!l!}(a^{2})^{l}(\zeta (-2\nu +1+2l,\frac{\beta }{2\pi })-(%
\frac{\beta }{2\pi })^{2\nu -1-2l})\\&\notag
+\sum_{k=1}^{\infty }\sum_{l=0}^{\sigma }\frac{(\nu -\frac{1}{2})!}{(\nu -\frac{1}{%
2}-l)!l!}(a^{2})^{l}(\zeta (-2\nu +1+2l,-\frac{\beta }{2\pi })-(-\frac{\beta }{%
2\pi })^{2\nu -1-2l})] .\end{align}

The objective now is to make the $\beta $ dependence clear. For
this we use the expansion of Hurwitz zeta \cite{qradstein7}:
\begin{equation}\label{176}
\zeta (z,q)=\frac{2\Gamma (1-z)}{(2\pi )^{1-z}}(\sin [\frac{\pi z}{2}%
]\sum_{n=1}^{\infty }\cos [\frac{2\pi qn}{n^{1-z}}]+\cos [\frac{\pi z}{2}%
]\sum_{n=1}^{\infty }\sin [\frac{2\pi qn}{n^{1-z}}])
.\end{equation} Also the $\zeta (z,-q)$ expansion, can be found
using \cite{kirstennew1},
\begin{equation}\label{177}
\zeta (1-s,a)=\frac{\Gamma (s)}{(2\pi )^{s}}(e^{-\frac{i\pi s}{2}}F(s,a)+e^{%
\frac{i\pi s}{2}}F(s,-a)) ,\end{equation} where
\begin{equation}\label{178}
F(s,a)=\sum_{n=1}^{\infty }\frac{e^{2i\pi na}}{n^{s}}
.\end{equation}%
which is valid if $Rez<0$ and $0<q\leq 1$

\noindent In our case $z=-2\nu +1+2l$. Note that for $d=3$, we
have $-2\nu =-4$ and $-2\nu +1+2l$ is negative for $l=0,1$. For
$l=2$ we use the Hurwitz zeta expansion, $\zeta (s,a)$, around
$s=1$, where a pole exists,
\begin{equation}\label{179}
\lim_{s\rightarrow 1}(\zeta (s,a)-\frac{1}{s-1})=-\psi _{0}(a)
.\end{equation}

\noindent Thus we can compute $V_{twist}$ as an expansion up to
order $L^{-2}$. By using dimensional regularization we Taylor
expand the $d$ dependent terms around $d+\varepsilon$,
$\varepsilon \rightarrow 0$ as before. Also for $d=3$ the
expression $-2\nu +1+2l$ is always an odd number for all $l$. So
the terms $(\frac{\beta}{2\pi })^{2\nu -1-2l}$ are omitted. Below
we quote the terms for $l=0,1,2$:
\begin{align}\label{180}
&V_{twist}=-\frac{1}{2}\frac{\sqrt{\pi }}{(2\pi )^{d}a}(2\pi )^{\frac{d-1}{2}%
}m^{d+1}\Gamma (-\nu -\frac{1}{2}+1)(\frac{\beta ^{2}}{a^{2}}+1)^{\nu +\frac{1}{2}%
-1}+\\&\notag \frac{1}{4}\frac{1}{(2\pi )^{d}}(2\pi
)^{\frac{d-1}{2}}m^{d+1}\Gamma (-\nu )\\&\notag
-\frac{1}{2}\frac{\sqrt{\pi }}{(2\pi )^{d}a}(2\pi )^{\frac{d-1}{2}%
}m^{d+1}\Gamma (-\nu -\frac{1}{2}+1)(a^{2})^{\frac{1}{2}-\nu }\ \\&\notag \times[((2\pi )^{2})^{\nu -%
\frac{1}{2}}\frac{2\Gamma (2\nu )}{(2\pi )^{2\nu }}\\&\notag \times(\sin [\frac{\pi (1-2\nu )}{2}%
]\sum_{n=1}^{\infty }\cos [\frac{\beta n}{n^{2\nu }}]+\cos [\frac{\pi (1-2\nu )}{2}%
]\sum_{n=1}^{\infty }\sin [\frac{\beta n}{n^{2\nu }}]\\&\notag
+\sin [\frac{\pi (1-2\nu )}{2}]\sum_{n=1}^{\infty }\cos [\frac{\beta n}{n^{2\nu }}%
]-\cos [\frac{\pi (1-2\nu )}{2}]\sum_{n=1}^{\infty }\sin
[\frac{\beta n}{n^{2\nu }}]\\&\notag
+\frac{(\nu -\frac{1}{2})!}{(\nu -\frac{1}{2}-1)!}a^{2}\\&\notag \times ((2\pi )^{2})^{\nu -\frac{1}{2}%
-1}[\frac{2\Gamma (2\nu -2)}{(2\pi )^{2\nu -2}}(\sin [\frac{\pi (3-2\nu )}{2}%
]\sum_{n=1}^{\infty }\cos [\frac{\beta n}{n^{2\nu -2}}]\\&\notag +\cos [\frac{\pi (3-2\nu )}{%
2}]\sum_{n=1}^{\infty }\sin [\frac{\beta n}{n^{2\nu -2}}]\\&\notag
+\sin [\frac{\pi (3-2\nu )}{2}]\sum_{n=1}^{\infty }\cos [\frac{\beta n}{n^{2\nu -2}%
}]-\cos [\frac{\pi (3-2\nu )}{2}]\sum_{n=1}^{\infty }\sin [\frac{\beta n}{%
n^{2\nu -2}}])\\&\notag
+\frac{(\nu -\frac{1}{2})!}{(\nu -\frac{1}{2}-2)!2!}a^{4}((2\pi )^{2})^{\nu -\frac{1}{%
2}-2}(\frac{2}{\varepsilon }+\psi _{o}(\frac{\beta }{2\pi })+\psi _{o}(-%
\frac{\beta }{2\pi })) ,\end{align}
\newpage
(with $\psi _{o}$ the digamma function) which after calculations
is written:
\begin{align}\label{181}
&V_{twist}=-\frac{1}{2}\frac{\sqrt{\pi }}{(2\pi )^{d}a}(2\pi )^{\frac{d-1}{2}%
}m^{d+1}\Gamma (-\nu -\frac{1}{2}+1)(\frac{\beta ^{2}}{a^{2}}+1)^{\nu +\frac{1}{2}%
-1}\\&\notag +\frac{1}{4}\frac{1}{(2\pi )^{d}}(2\pi
)^{\frac{d-1}{2}}m^{d+1}\Gamma (-\nu )\\&\notag
-\frac{1}{2}\frac{\sqrt{\pi }}{(2\pi )^{d}a}(2\pi )^{\frac{d-1}{2}%
}m^{d+1}\Gamma (-\nu -\frac{1}{2}+1)(a^{2})^{\frac{1}{2}-\nu }{\LARGE (}\frac{%
2\Gamma (2\nu )}{(2\pi )^{2\nu }}((2\pi )^{2})^{\nu -\frac{1}{2}}\\&\notag \times{\Large [}%
{\normalsize 2}\sin (\frac{\pi (1-2\nu )}{2})\sum_{n=1}^{\infty }\cos (\frac{%
\beta n}{n^{2\nu }})){\Large ]\ }\\&\notag
+\frac{(\nu -\frac{1}{2})!}{(\nu -\frac{1}{2}-1)!}a^{2}((2\pi )^{2})^{\nu -\frac{1}{2}%
-1}{\Large [}\frac{2\Gamma (2\nu -2)}{(2\pi )^{2\nu -2}}(2\sin (\frac{\pi (3-2\nu )}{2%
})\sum_{n=1}^{\infty }\cos (\frac{\beta n}{n^{2\nu -2}})\\&\notag
+\frac{(\nu -\frac{1}{2})!}{(\nu -\frac{1}{2}-2)!2!}a^{4}((2\pi )^{2})^{\nu -\frac{1}{%
2}-2}{\Large [}\frac{2}{\varepsilon }+\psi _{o}(\frac{\beta }{2\pi
})+\psi _{o}(-\frac{\beta }{2\pi }){\Large ]}{\LARGE
)}+O(\varepsilon ,\varepsilon ^{2}\mathrm{and\ higher}))
,\end{align} with $\beta =2\pi \omega ,\ \nu =\frac{d+1}{2},\
a=mL\ $. The sums appearing above are:
\begin{equation}\label{182}
\sum_{n=1}^{\infty }\cos (\frac{\beta n}{n^{2\nu }})=\frac{1}{2}%
(Li_{2\nu }(e^{-i\beta })+Li_{2\nu }(e^{i\beta })) ,\end{equation}
and
\begin{equation}\label{183}
\sum_{n=1}^{\infty }\cos (\frac{\beta n}{n^{2\nu -2}})=\frac{1}{2}%
(Li_{2\nu -2}(e^{-i\beta })+Li_{2\nu -2}(e^{i\beta }))
.\end{equation} Let us see how the poles cancel in the above
expressions. In the case $d=3$ one of the poles is contained to
the Hurwitz, and is of the form $\frac{2}{\varepsilon }$ with
$\varepsilon \rightarrow 0$. The other pole is contained to the expression $\frac{1}{4}\frac{1}{(2\pi )^{d}}%
(2\pi )^{\frac{d-1}{2}}m^{d+1}\Gamma (-\nu )$. Thus we have:
\begin{align}\label{184}
&V_{twist}=\frac{\frac{-m^{4}}{16\,\pi ^{2}}+\frac{m^{4}}{%
16\,\pi ^{2}}}{\varepsilon }+(\frac{3\,m^{4}}{64\,\pi ^{2}}-\frac{%
\gamma \,m^{4}}{32\,\pi ^{2}}\\&\notag
\frac{m^{4}\,\left( 1+\frac{\beta ^{2}}{\alpha ^{2}}\right) ^{\frac{3}{2}}%
}{6\,\pi \,\alpha }+\frac{2\,m^{4}\,\sqrt{\alpha ^{2}}\,\cos
(\beta )}{\pi ^{2}\,\alpha ^{5}}+\frac{m^{4}\,\ln (2)}{16\,\pi
^{2}}\\&\notag +\frac{m^{4}\,\ln (\pi )}{32\,\pi
^{2}}+\frac{m^{4}\,\ln (\pi )}{32\,\pi ^{2}\,}\\&\notag
-\frac{m^{4}\,\ln (\alpha ^{2})}{32\,\pi ^{2}\,}-\frac{m^{4}
\,\psi (-\frac{3}{2})}{32\,\pi ^{2}\,}-\frac{%
m^{4}\,\psi (\frac{1}{2})}{32\,\pi ^{2}\,}
\\&\notag +\frac{m^{4}\,\psi
(\frac{5}{2})}{32\,\pi
^{2}\,}+\frac{m^{4}\,\psi (\frac{-\beta }{%
2\,\pi })}{32\,\pi ^{2}\,}+\frac{m^{4}\,\psi (\frac{\beta }{2\,\pi
})}{32\,\pi ^{2}\,}) .\end{align} We can see how the poles cancel.
The last expression is the vacuum energy in the case that
arbitrary phases appear.

\subsection{Some Applications II}

\subsubsection{Extra Dimensional Models with Twisted Boundary Conditions}

Let us now briefly present an application of the twisted potential
case we computed above.

\noindent In models with large extra dimensions, supersymmetry can
be broken in the bulk by the Scherk-Schwarz mechanism, as we
described briefly in the introduction. Consider the immediate
extra dimensional extension of the MSSM in five dimensions on the
orbifold $S^1/Z_2$ \cite{antoniadis,antoniadis123,extra1,extra2}.
Assume that supersymmetry breaking occurs in the bulk through the
Scherk-Schwarz mechanism \cite{scherk}. Thus the fields have the
following boundary conditions,
\begin{equation}\label{boundarspygame}
\Phi (x^{\mu},y+2\pi R)=e^{2\pi iq_{\Phi}}\Phi
(x^{\mu},y).\end{equation} The Scherk-Schwarz mechanism consists
in using different parameters $q_{\Phi}$ for fermions and bosons
belonging to the same hypermultiplet. The harmonic expansion of
the fields for circle compactification is,
\begin{equation}
\Phi(x^{\mu},y)=\sum_{n=-\infty }^{\infty
}{\Phi}_{n}(x)e^{\frac{i2{\pi}(n+q_{\Phi})y}{R}}.\end{equation} In
the case of the $S^1/Z_2$ orbifold compactification, the $Z_2$
even fields have harmonic expansion,
\begin{equation}
\Phi(x^{\mu},y)=\sum_{n=-\infty }^{\infty
}{\Phi}_{n}(x)\cos{\frac{2{\pi}(n+q_{\Phi})y}{R}},\end{equation}
while the $Z_2$ odd fields,
\begin{equation}
\Phi(x^{\mu},y)=\sum_{n=-\infty }^{\infty
}{\Phi}_{n}(x)\sin{\frac{2{\pi}(n+q_{\Phi})y}{R}}.\end{equation}
The $Z_2$ even fields have zero modes and produce the 4
dimensional MSSM, while the $Z_2$ odd don't have zero modes. The
Kaluza-Klein modes within each hypermultiplet have masses,
\begin{equation}\label{bosonmmssm}
m_B^2=\frac{(n+q_B)^2}{R^2},\end{equation} for the boson case, and
for the fermion case the mass reads,
\begin{equation}\label{bosonmmssm}
m_F^2=\frac{(n+q_F)^2}{R^2}.\end{equation} In the orbifold extra
dimensional extension, the electroweak symmetry breaking occurs
through radiative corrections to the Higgs mass. So it is
necessary to include one loop corrections to the appropriate mass
eigenstate Higgs scalar field mass (for more details see
\cite{extra2,extra1}). The one loop corrected mass is induced by a
tower of KK states and is equal to,
\begin{equation}\label{masscorrection}
m_{\phi}^2(\phi=0)=\frac{\mathrm{d}^2V(\phi)}{\mathrm{d}\phi^2},\end{equation}
with $V(\phi)$ given by,
\begin{equation}\label{extrdimensmoglied}
V(\phi)=\frac{1}{2}\mathrm{Tr}\sum_{n=-\infty }^{\infty }\int
\frac{\mathrm{d^4p}}{(2\pi)^4}\ln
\Big{[}\frac{p^2+\frac{(n+q_B)^2}{R^2}+M^2(\phi)}{{p^2+\frac{(n+q_F)^2}{R^2}+M^2(\phi)}}\Big{]}.\end{equation}
In the above, $M^2(\phi)$ is the $\phi$-dependent mass of the KK
states which are model dependent. It is obvious that the effective
potential (\ref{extrdimensmoglied}) is identical to (\ref{121})
which was computed in the previous section. Thus the
Scherk-Schwarz phases are like twists in the boundary conditions.
The calculation follows as we described above. See also
\cite{kirstennew1,elizalde}.

\subsection{An Alternative Elegant Approach. Epstein Zeta Functions}

In this section we briefly present a much more elegant and more
elegant computation method for the effective potential. Consider a
massive scalar field quantized in $T^N\times R^n$ with periodic
boundary conditions in each of the torii, that is,
\begin{equation}\label{tntorii}
\phi (x_i)=\phi (x_i+L_i) ,\end{equation} with $x_i$ the
coordinates describing the torii and $L_i$ the torii radii. The
zeta function corresponding to this setup is
\cite{kirstennew1,elizalde,elizalde2,kirstennew3,kirstennew5,elizaldekirstenarme,kirstenepstein},
\begin{equation}\label{tombishp}
\zeta (s,L_i)=(2\pi)^{-n}\sum_{n_1...n_N=-\infty}^{\infty}\int
\mathrm{d}^nk\Big{[}\Big{(}\frac{2\pi
n_1}{L_1}\Big{)}+...+\Big{(}\frac{2\pi
n_N}{L_N}\Big{)}+k^2+M^2\Big{]}^{-s} ,\end{equation} The general
summations can be written in terms of the Epstein zeta function.
Indeed after performing the integration in relation
(\ref{tombishp}), we obtain,
\begin{equation}\label{bishoptom}
\zeta
(s,w_i)=\Big{(}\frac{\sqrt{\pi}}{L_1}\Big{)}^n\frac{\Gamma(s-n/2)}{\Gamma
(s)}\Big{(}\frac{L_1}{2\pi}\Big{)}^{2s}Z_N^{v^2}\Big{(}s-n/2;w_1,...,w_N\Big{)}
,\end{equation} with $w_i=(L_1/L_i)^2$. In the above we used the
generalized Epstein zeta function,
\begin{equation}\label{cativoepsteinzeta}
Z_N^{v^2}\Big{(}s;w_1,...,w_N\Big{)}=\sum_{n_1...n_N=-\infty}^{\infty}[w_1n_1^2+...+w_Nn_N^2+v^2]^{-s}.\end{equation}

\noindent The interested reader can consult the references
\cite{elizaldekirstenarme,kirstenepstein,elizalde,kirstennew1},
where the subject is developed in great detail.

\subsection{Twisted Sections and Non Trivial Topology}

\noindent One question that one might ask is if there a criterion
or more correctly a way to know which are the allowed boundary
conditions for a field in a specific topology. The answer can be
given in terms of the allowed sections of the fibre bundles that
the spacetime topology corresponds to.

\noindent Non trivial topology affects the fields entering the
Lagrangian (twisted fields) (see for example
\cite{isham,gongcharov,ford,spalluci}).\ In our case, the
topological properties of $S^{1}{\times R}^{3}$ are classified by
the first Stieffel class $H^{1}(S^{1}{\times
R}^{3},Z_{\widetilde{2}})$ which is isomorphic to the singular
(simplicial) cohomology group ${H}_{1}({S} ^{1}{\times
R}^{3}{,Z}_{2})$ because of the triviality of the
${Z}_{\widetilde{2}}$ sheaf. It is known that
$H^{1}{(S}^{1}{\times R}^{3}{,Z}_{\widetilde{2}}{)=Z}_{2}$
classifies the twisting of a bundle. Specifically, it describes
and classifies the orientability of a bundle globally. In our
case, the classification group is ${Z}_{2}$ and, we have two
locally equivalent bundles, which are however different globally
(like in the case of the cylinder and that of the moebius strip
where both locally resemble $S^1\times R$). The mathematical lying
behind, is to find the sections that correspond to these two fibre
bundles, and which are classified by $Z_{2}$ \cite{isham}. The
sections we used are real scalar fields and Majorana or Dirac
spinor fields. These carry a topological number called moebiosity
(twist), which distinguishes between twisted and untwisted fields.
The twisted fields obey anti-periodic boundary conditions, while
untwisted fields periodic in the compact dimension. In the finite
temperature case one takes scalar fields to obey periodic and
fermion fields anti-periodic boundary conditions, disregarding all
other configurations that may arise from non trivial topology. We
shall consider all these configurations. Let $\varphi _{u}$,
$\varphi_{t}$ and $\psi _{t}$, $\psi _{u}$ denote the untwisted
and twisted scalar and twisted and untwisted spinor fields
respectively. The boundary conditions in the ${S}^{1}$ dimension
read,
\begin{equation}\label{bc1}
\varphi_{u}(x,0)=\varphi _{u}(x,L),\end{equation} and
\begin{equation}\label{bc2}
\varphi _{t}(x,0)=-\varphi _{t}(x,L),\end{equation} for scalar
fields and
\begin{equation}\label{bc1}
\psi _{u}(x,0)=\psi _{u}(x,L),\end{equation} and
\begin{equation}\label{bc2}
\psi_{t}(x,0)=-\psi _{t}(x,L),\end{equation} for fermion fields,
where $x$ stands for the remaining two spatial and one time
dimension which are not affected by the boundary conditions.
Spinors (both Dirac and Majorana), still remain Grassmann
quantities. The untwisted fields are assigned twist ${h}_{0}$ (the
trivial element of ${Z}_{2})$ and the twisted fields twist $h_{1}$
(the non
trivial element of ${Z}_{2}$).\ Recall that $h_{0}+h_{0}=h_{0}$ ($0+0=0$), $%
h_{1}+h_{1}=h_{0}$ ($1+1=0$), $h_{1}+h_{0}=h_{1}$ ($1+0=1$). We
require the Lagrangian to be scalar under $Z_{2}$ thus to have
${h}_{0}$ moebiosity. Thus the
topological charges flowing at the interaction vertices must sum to ${h}_{0}$ under ${H}^{1}{(S}^{1}{\times R}^{3}{,Z%
}_{\widetilde{2}}{)}$. For supersymmetric models, supersymmetry
transformations impose some restrictions on the twist assignments
of the superfield component fields \cite{gongcharov}.

\noindent No other field configuration is allowed to take non zero
vev but the untwisted scalars. This is due to Grassmann nature of
the vacuum or space dependent vacuum solutions that other
configurations imply.

\noindent In the general case when the spacetime has topology
$(S^1)^q\times R^{4-q}$, then the topologically allowed field
configurations are classified by the representations of
$H^1\big{(}(S^1)^q\times R^{4-q},Z_2\big{)}=Z_2^q$. Thus the
different inequivalent twists that can be assigned are $2^q$. This
means that we can have $2^q$ topologically inequivalent spin $0$
scalars, spin $1/2$ Majorana fermions and spin $3/2$ Majorana
fermions (this for supergravity). For our case $q=1$.

\noindent It worths mentioning equivalent mathematical setups that
exist in the literature. Twisted fields have frequently been used,
for example as we seen in the Scherk-Schwarz mechanism
\cite{scherk} for supersymmetry breaking in our 4-dimensional
world, where the harmonic expansion of the fields is of the form:
\begin{equation}
\phi(x,y)=e^{imy}\sum_{n=-\infty }^{\infty
}{\phi}_{n}(x)e^{\frac{i2{\pi}ny}{L}},\end{equation} The ''$m$''
parameter incorporates the twist mentioned above. This treatment
is closely related to automorphic field theory \cite{Dowker} in
more than 4 dimensions (which is an alternative to the one used by
us).

\noindent Concerning the automorphic field theory, due to the
compact dimension we can use generic boundary conditions for
bosons and fermions in the compact
dimension which are,%
\begin{eqnarray}
\varphi _{i}(x_{2},x_{3},\tau ,x_{1}) &=&e^{i\pi n_{1}\alpha
}\varphi
_{i}(x_{2},x_{3},\tau ,x_{1}+L) \\
~\Psi (x_{2},x_{3},\tau ,x_{1}) &=&e^{i\pi n_{1}\delta }\Psi
(x_{2},x_{3},\tau ,x_{1}+L),  \notag
\end{eqnarray}%
with, $0<\alpha ,\delta <1$, $i=1,2$, $n_{1}=1,2,3...$. The values
$\alpha= 0,1$ correspond to periodic and antiperiodic bosons
respectively while $\delta =0,1$ corresponds to periodic and
anti-periodic fermions \cite{Dowker}.

\subsection{The Validity of Approximations. Numerical Tests}

Let us check numerically one of our results. We focus on the
bosonic contribution at high temperature.
\begin{figure}[h]
\begin{center}
\includegraphics[scale=.7]{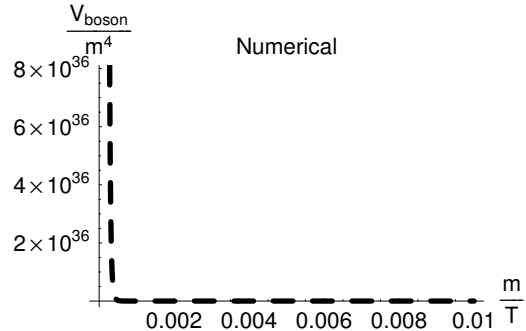}
\end{center}
\caption{Plot of the dependence of $V_{boson}/m^{d+1}$ as a
function of $m/T$. Numerical approximation of Bessel sum.
5-dimensional bosonic theory at finite temperature.}
\label{figure1}
\end{figure}
\begin{figure}[h]
\begin{center}
\includegraphics[scale=.7]{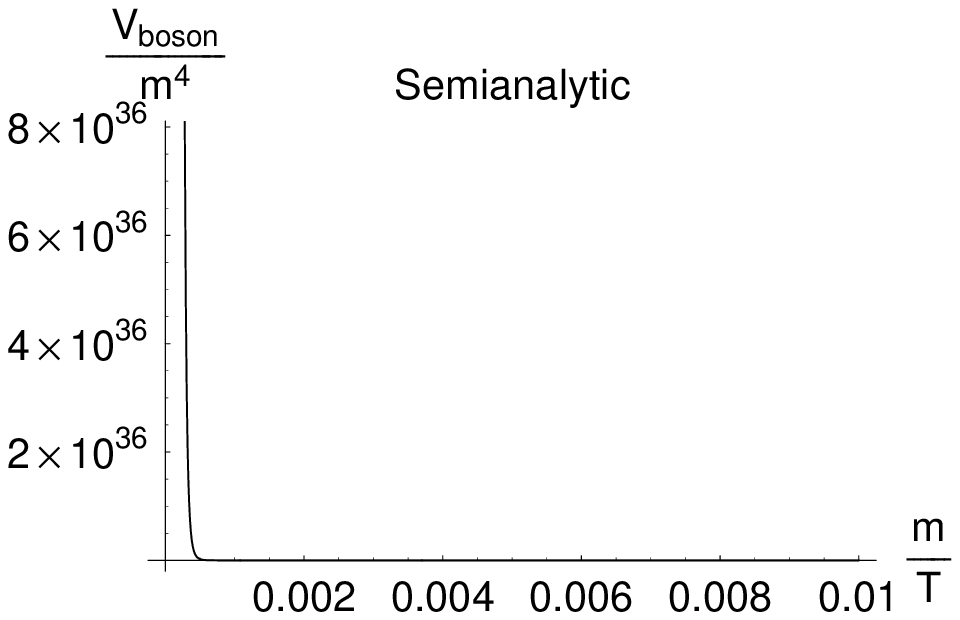}
\end{center}
\caption{Plot of the dependence of $V_{boson}/m^{d+1}$ as a
function of $m/T$. Semi-analytic approximation. 5-dimensional
bosonic theory at finite temperature.} \label{figure2}
\end{figure}
\begin{figure}[h]
\begin{center}
\includegraphics[scale=.9]{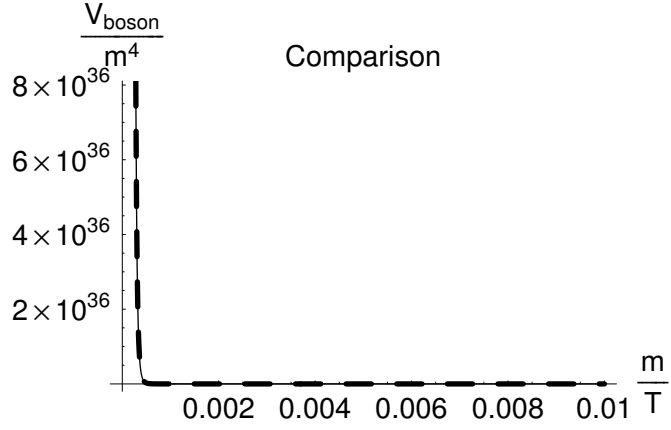}
\end{center}
\caption{Comparison of numerical and corresponding semi-analytic
approximation. 5-dimensional bosonic theory at finite
temperature.} \label{figure3}
\end{figure}
We shall study the convergence properties of our approximation and
how the semi-analytic results behave in comparison to the
numerical evaluation of the potential. As we seen, before the high
temperature limit was taken, the bosonic contribution is given by:
\begin{equation}\label{185}
V_{boson}=-\sum\limits_{q=1}^{\infty }\frac{1}{(2\pi )^{d}}(2\pi )^{\frac{d-1}{%
2}}m^{d+1}\frac{K_{\frac{d+1}{2}}(\frac{mq}{T})}{(\frac{mq}{2T})^{\frac{d+1}{2}}}
.\end{equation} \noindent After the high temperature limit was
taken, the effective potential is given by the semi-analytic
approximation:
\begin{align}\label{188}
&V_{boson}=-\frac{1}{2}\frac{\sqrt{\pi }}{(2\pi )^{d}a}(2\pi )^{\frac{d-1}{2}%
}m^{d+1}\Gamma (-\nu -\frac{1}{2}+1)\\&\notag
+\frac{1}{4}\frac{1}{(2\pi )^{d}}(2\pi
)^{\frac{d-1}{2}}m^{d+1}\Gamma (-\nu )\\&\notag
-\frac{\sqrt{\pi }}{(2\pi )^{d}a}(2\pi )^{\frac{d-1}{2}%
}m^{d+1}\Gamma (-\nu -\frac{1}{2}+1)(a^{2})^{\frac{1}{2}-\nu }\\&\notag \times [\sum_{l=0}^{\sigma }%
\frac{((2\pi )^{2})^{\nu -\frac{1}{2}-l}(\nu -\frac{1}{2})!}{(\nu -\frac{1}{2}-l)!l!}%
(a^{2})^{l}\zeta (-2\nu  +1+2l)].
\end{align}
The converge of (\ref{188}) and (\ref{185}) is quite fast. Also
the two relations describe the same physics and are identical as
can be checked. Particularly this holds even if we keep only a few
terms of (\ref{188}). We have checked this for values of $m/T$
that our approximation is valid, that is $m/T<1$. Also this holds
for several dimensions. Let us study the finite temperature limit
of a 5-dimensional theory, that is for $d=4$. In figure
\ref{figure1} we plot the dependence of $V_{boson}/m^{d+1}$ as a
function of $m/T$, where $V_{boson}$ is given by the Bessel sum of
relation (\ref{185}). A numerical calculation is done for the sum
over the Bessel functions. Also in figure \ref{figure2} we plot
the dependence of $V_{boson}/m^{d+1}$ as a function of $m/T$, with
$V_{boson}$ given by the semi-analytic approximation of relation
(\ref{188}). In addition, in figure \ref{figure3} we compare the
above results. As we can see the two results are identical for a
large range of the expansion parameter $m/T$.
\begin{figure}[h]
\begin{center}
\includegraphics[scale=.7]{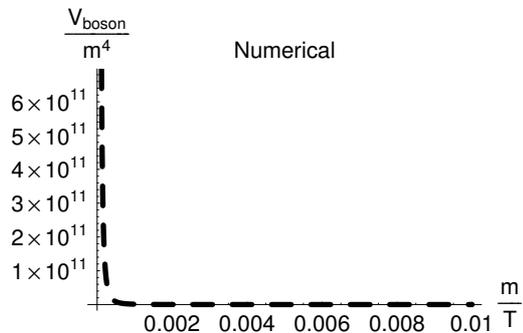}
\end{center}
\caption{Plot of the dependence of $V_{boson}/m^{d+1}$ as a
function of $m/T$. Numerical approximation of Bessel sum.
4-dimensional bosonic theory at finite temperature.}
\label{figure4}
\end{figure}
\begin{figure}[h]
\begin{center}
\includegraphics[scale=.7]{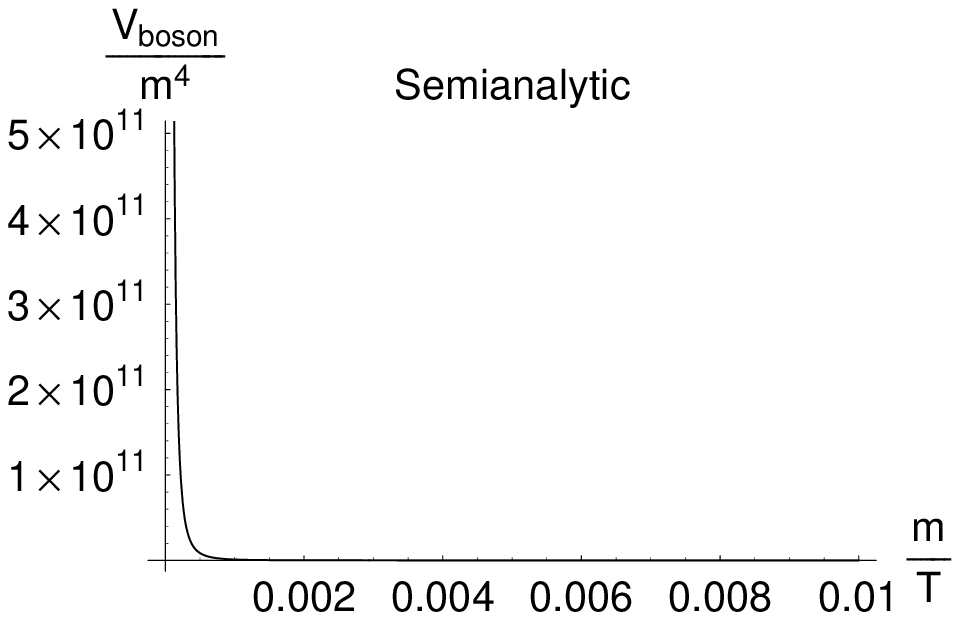}
\end{center}
\caption{Plot of the dependence of $V_{boson}/m^{d+1}$ as a
function of $m/T$. Semi-analytic approximation. 4-dimensional
bosonic theory at finite temperature.} \label{figure5}
\end{figure}
\begin{figure}[h]
\begin{center}
\includegraphics[scale=.9]{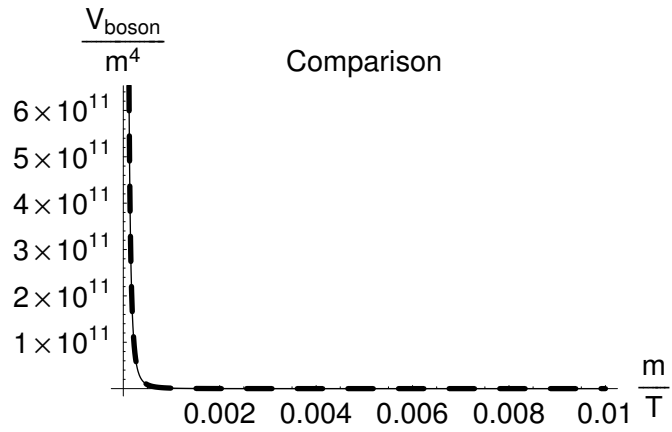}
\end{center}
\caption{Comparison of numerical and corresponding semi-analytic
approximation. 4-dimensional bosonic theory at finite
temperature.} \label{figure6}
\end{figure}
This shows us that in the high temperature limit ($\frac{m}{T}<1$)
the semi-analytic expressions we obtained are in complete
agreement to the numerical values. This holds regardless the
number of terms of the semi-analytic expansion we keep. Thus the
expansion is perturbative and valid. The same analysis can be done
for the $d=4$ case. We present the results in figures
\ref{figure4}, \ref{figure5} and \ref{figure6}. Thus within the
perturbative limits the semi-analytic approximation is valid and
exponentially converging as expected (see also \cite{elizalde}).

\newpage

\section*{Acknowledgements}

\noindent The author would like to thank the referee of Reviews in
Mathematical Physics for invaluable comments and suggestions that
improved significantly the quality and appearance of the paper.

\end{document}